\documentclass[twocolumn, 
]{aastex7}
\usepackage{nccmath,amsmath}
\usepackage{booktabs}
\usepackage{siunitx}
\usepackage{caption}
\usepackage{graphicx} 		
\usepackage{subfigure}
\usepackage[retainorgcmds]{IEEEtrantools}

\newcommand{\ud}{\mathrm{d}}

\defcitealias{2020A&A...641A...6P}{\emph{Planck-2018}}

\shorttitle{Deep learning emulator for inflationary GWB}
\shortauthors{Zhang et al.}
\graphicspath{{./}{figs/}}

\begin{document}

\title{SageNet: Fast Neural Network Emulation of the Stiff-amplified Gravitational Waves from Inflation}
\correspondingauthor{Bohua Li}

\author[orcid=0009-0004-3509-7495,sname='Zhang']{Fan Zhang}
\affiliation{Zhejiang University, State Key Laboratory of Ocean Sensing \& Ocean College, Zhoushan, Zhejiang 316000, China}
\affiliation{Massachusetts Institute of Technology, Kavli Institute for Astrophysics and Space Research, Cambridge, MA, 02139, USA}
\email{f.zhang@zju.edu.cn}  

\author[orcid=0009-0006-7175-5373,gname=Yifang, sname=Luo]{Yifang Luo}
\affiliation{Zhejiang University, State Key Laboratory of Ocean Sensing \& Ocean College, Zhoushan, Zhejiang 316000, China}
\email{luoyifang@zju.edu.cn}

\author[orcid=0000-0002-3600-0358,gname=Bohua, sname=Li]{Bohua Li}
\affiliation{Guangxi Key Laboratory for Relativistic Astrophysics, 
School of Physical Science and Technology, Guangxi University \\
Nanning, Guangxi, 530004, China}
\email[show]{bohuali@gxu.edu.cn}

\author[gname=Ruihan,sname=Cao]{Ruihan Cao}
\affiliation{Newton South High School, 140 Brandeis Road, Newton Centre, MA 02459, USA}
\email{ruihan20080129@gmail.com}

\author[gname=Wenjin,sname=Peng]{Wenjin Peng}
\affiliation{Wuhan University, School of Cyber Science and Engineering, Wuhan, Hubei, 430072, China}
\email{peng.wenjin@whu.edu.cn}

\author[orcid=0000-0001-8510-2812]{Joel Meyers}
\affiliation{Department of Physics, Southern Methodist University, Dallas, TX 75275, USA}
\email{jrmeyers@mail.smu.edu}

\author[orcid=0000-0002-0410-3045,gname=Paul,sname=Shapiro]{Paul R. Shapiro}
\affiliation{Department of Astronomy, The University of Texas at Austin, Austin, TX 78712, USA}
\email{shapiro@astro.as.utexas.edu}



\begin{abstract}

Accurate modeling of the inflationary gravitational waves (GWs) requires time-consuming, iterative numerical integrations of differential equations to take into account their backreaction on the expansion history. To improve computational efficiency while preserving accuracy, we present the Stiff-Amplified Gravitational-wave Emulator Network (\verb|SageNet|), a deep learning framework designed to replace conventional numerical solvers\footnote{Code available at: \url{https://github.com/YifangLuo/SageNet}}. \verb|SageNet| employs a long short-term memory architecture to emulate the present-day energy density spectrum of the inflationary GWs with possible stiff amplification, $\Omega_\mathrm{GW}(f)$. Trained on a data set of 25,689 numerically generated solutions, \verb|SageNet| allows accurate reconstructions of $\Omega_\mathrm{GW}(f)$ and generalizes well to a wide range of cosmological parameters; 90.9\% of the test emulations with randomly distributed parameters exhibit errors of under 4\%. In addition, \verb|SageNet| demonstrates its ability to learn and reproduce the artificial, adaptive sampling patterns in numerical calculations, which implement denser sampling of frequencies around changes of spectral indices in $\Omega_\mathrm{GW}(f)$. The dual capability of learning both physical and artificial features of the numerical GW spectra establishes \verb|SageNet| as a robust alternative to exact numerical methods. Finally, our benchmark tests show that \verb|SageNet| reduces the computation time from tens of seconds to milliseconds, achieving a speed-up of $\sim 10^4$ times over standard CPU-based numerical solvers with the potential for further acceleration on GPU hardware. These capabilities make \verb|SageNet| a powerful tool for accelerating Bayesian inference procedures for extended cosmological models. In a broad sense, the \verb|SageNet| framework offers a fast, accurate, and generalizable solution to modeling cosmological observables whose theoretical predictions demand costly differential equation solvers.

\end{abstract}



\section{Introduction}\label{sec:intro}

The stochastic gravitational-wave background (SGWB)
from primordial tensor fluctuations
is an important prediction of the inflationary paradigm
\citep{1979JETPL..30..682S,1982PhLB..115..189R,1984NuPhB.244..541A}. 
If the primordial tensor power spectrum is allowed to be blue-tilted
(i.e., of positive tensor spectral index $n_\mathrm{t}$),
the inflationary SGWB may be directly measured by a combination of ongoing observations,
including the cosmic microwave background (CMB) polarization experiments
\citep{1997PhRvL..78.2054S,1997PhRvL..78.2058K},
pulsar timing arrays (PTAs) \citep{1983ApJ...265L..39H}
and laser interferometer gravitational-wave (GW) experiments
\citep[e.g.,][]{2016PhRvX...6a1035L, 2018JCAP...11..038K, 2018CQGra..35p3001C},
thanks to the long lever arm spanned by the scales probed by these observations
\citep{2006PhRvD..73l3503S,2015PhRvD..91j3505M}.
In fact, several PTA collaborations have recently reported
strong evidence for an SGWB in the nanohertz frequency band
\citep{2023RAA....23g5024X, 2023ApJ...951L...8A, 2023A&A...678A..50E, 2023ApJ...951L...6R}
and the blue-tilted inflationary SGWB interpretation
is consistent with the current PTA data
\citep{2023ApJ...951L..11A, 2024A&A...685A..94E,2024PhRvL.132q1002F,2025ApJ...985..117L}.

In the meantime, the blue-tilted inflationary SGWB can also influence
the energy budget of the Universe and hence the expansion history significantly
in the radiation-dominated (RD) era
\citep[e.g.,][]{1969JETPL...9..184S,2000PhR...331..283M,2008PhRvD..78d3531B,2017PhRvD..96f3505L}.
This contribution is conventionally described by
the effective number of \emph{extra} relativistic species, $\Delta N_\mathrm{eff}$,
which provides an indirect probe of relic GWs produced in the early Universe.
\citet{2021JCAP...10..024L} have shown that
an inflationary SGWB with $\Delta N_\mathrm{eff,GW}\sim 0.3$
can cause a percent-level shift of the expansion rate during the RD era.
Nonetheless, this backreaction of the SGWB on the expansion history
\citep{1998PhRvD..58h3504G}
is often overlooked in existing Bayesian analyses of GW data sets,
leading to possible errors in the inferred cosmological parameters.
In order to model the present-day inflationary SGWB spectrum accurately, 
self-consistent treatments of the backreaction
typically involve an iterative algorithm \citep{2022MNRAS.509.1366K},
since the evolution of each tensor mode is itself dependent on the expansion history.

Furthermore, additional inaccuracies in the prediction
of the SGWB energy density spectrum today, $\Omega_\mathrm{GW}(f)$,
can arise from the treatment of the tensor transfer function, $T(f)$.
Since the tensor wave equation that governs the evolution of tensor modes
does not have analytical solutions for modes that reentered the horizon
when the equation of state (EoS) of the Universe varies,
exact integration of the wave equation is thus required for these modes.
This approach results in more accurate predictions for $T(f)$
than the usual approach of using a fitting formula,
which often neglects the backreaction effect
\citep[e.g.,][]{1993PhRvD..48.4613T,2015JCAP...02..003K}.

Therefore, accurate physical modeling of the inflationary SGWB should
take into account both the exact evolution of tensor modes
and the backreaction effect.
In previous work, some of us have incorporated these elements
in the \verb|stiffGWpy| code (link provided at the end of the paper),
which solves the tensor wave equation for a range of adaptively chosen frequencies
to capture the entire shape of $\Omega_\mathrm{GW}(f)$.
Note that although the cosmological model considered in the \verb|stiffGWpy|
code allows for the possibility of an early kination/stiff phase
(see Section~\ref{sec:SGWB} for a complete description),
the presence of a stiff phase is not an essential part of the model.
Based on this extended model, 
\citet{2025ApJ...985..117L} performed Bayesian fit analyses
on current and mock PTA data,
using either nested sampling or Markov Chain Monte Carlo (MCMC) sampling.

While the above PTA analyses yielded correct posterior probability distributions
for the model parameters, the computational efficiency was however poor,
mostly due to the bottleneck in generating theoretical predictions
by iteratively solving ordinary differential equations (ODEs)
($\sim 10$\,s for each sample).
Meanwhile, only a subset of the free parameters of the \verb|stiffGWpy| code
were sampled in \citet{2025ApJ...985..117L};
cosmological parameters that are not directly related to the inflationary SGWB,
e.g., $(h,\,\Omega_\mathrm{m}h^2,\,A_\mathrm{s})$, were held fixed.
As the dimension of the parameter space increases,
the inefficiency of the theory code will pose a serious challenge
to Bayesian fit analyses based on the full-scale inflationary SGWB model.
The computational bottleneck in generating the theoretical $\Omega_\mathrm{GW}(f)$
would practically prohibit the completion of any sampling algorithm over the multidimensional parameter space.
Efficient sampling can only be realized
if the calculation of $\Omega_\mathrm{GW}(f)$ is accelerated.

In this paper, we tackle the above task by deep learning acceleration.
Deep learning has recently emerged as a promising alternative
to existing numerical processes, 
leveraging its robust nonlinear modeling capabilities
and hardware acceleration potential.
For instance, \citet{marx_machine-learning_2024} trained a neural network
with GW signal samples generated by the \verb|Bilby| \citep{ashton_bilby_2019}
and \verb|IMRPhenomPv2| \citep{hannam_simple_2014} libraries,
achieving accurate GW searches with latency on the order of seconds.
Based on this GW search pipeline, 
\citet{chatterjee_rapid_2024} developed a real-time parameter estimation algorithm
using an embedding network, with delays still constrained to seconds.
Similarly, \citet{raikman_gwak_2024} employed data
obtained by comparable methods to train a long short-term memory (LSTM) autoencoder
within a semi-supervised framework,
identifying deviations from normal GW patterns.

In addition, by establishing data-driven models, deep learning
can approximate functional relationships in computational tasks
by neural network-based nonlinear transformations,
bypassing conventional numerical schemes.
This approach is particularly useful for astrophysical processes
described by stable, deterministic physics. 
\citet{derose_neural_2022} utilized fully connected neural networks
to replace the Boltzmann solver,
accelerating the calculation of power spectra
required for the analysis of galaxy clustering and weak lensing data. 
When the data has a sequential or iterative nature,
using a recurrent neural network \citep{escamilla-rivera_deep_2020}
or its variant, LSTM \citep{bai_predicting_2021},
can reach higher accuracies.
\citet{yan_modeling_2024} demonstrated that
LSTM can efficiently find solutions
to multidimensional partial differential equations
even without knowing the specific form of the equations,
thereby attaining faster modeling of the time evolution of compact binary systems
than conventional numerical methods.


We herein develop the Stiff-amplified Gravitational-wave Emulator Network (\verb|SageNet|), an LSTM-based network
that can accurately reconstruct the asymptotic solutions to the tensor wave equation.
The resultant emulator can efficiently produce
the $\Omega_\mathrm{GW}(f)$ of the inflationary SGWB
for arbitrary model parameters within reasonable prior ranges.
The rest of the paper is organized as follows.
In Section~\ref{sec:SGWB}, we present our physical model
of the inflationary SGWB with possible stiff amplification.
In Section~\ref{sec:design}, we propose a deep learning-based emulator
for the inflationary SGWB and discuss its design.
In Section~\ref{sec:network},
we describe our neural network model and the training process.
We provide a comprehensive performance analysis
of the emulator in Section~\ref{sec:performance}
and conclude in Section~\ref{sec:conclusion}.



\section{Physical Model: Stiff-amplified Inflationary SGWB}\label{sec:SGWB}

The description of our physical model of the inflationary SGWB
closely follows \citet{2021JCAP...10..024L}.
The primordial tensor power spectrum satisfies a power law:
$\Delta_\mathrm{t}^2(f)=A_\mathrm{t}\,(f/f_\mathrm{CMB})^{n_\mathrm{t}}$, 
where the tensor amplitude $A_\mathrm{t}$ is related to the scalar amplitude
by the tensor-to-scalar ratio, $r\equiv A_\mathrm{t}/A_\mathrm{s}$, 
and $k_\mathrm{CMB}\equiv 2\pi f_\mathrm{CMB}/c=0.05\,\text{Mpc}^{-1}$
is the CMB pivot scale \citep{2020A&A...641A..10P}. 
Both $r$ and $n_\mathrm{t}$ (tensor spectral index) are free parameters of the model.

On top of the primordial tensor spectrum,
we consider the effect that
$\Omega_\mathrm{GW}(f)$ can be additionally blue-tilted
in the presence of a \emph{kination} phase in the early expansion history,
in which the Universe is dominated by the kinetic energy of some scalar field
\citep{1993PhLB..315...40S, 1997PhRvD..55.1875J, 2020PhRvL.124y1802C}.
Kination is also known as the ``stiff phase,''
since the EoS during kination is that of a stiff fluid,
i.e. $w\equiv\bar P/\bar\rho=1$ \citep{2014PhRvD..89h3536L, 2025SCPMA..6880409Y}.
When kination is present, the tensor modes that reentered the horizon during kination
shall end up in the SGWB with $\Omega_\mathrm{GW}(f)\propto f^{n_\mathrm{t}+1}$
\citep{1998PhRvD..58h3504G,2008PhLB..668...44G,2008PhRvD..77f3504B,2011PhRvD..84l3513K},
instead of $\Omega_\mathrm{GW}(f)\propto f^{n_\mathrm{t}}$
as for modes reentered during the RD era ($w_\mathrm{RD}=1/3$).
This effect is called the kination/stiff amplification of the inflationary SGWB
\citep{2017PhRvD..96f3505L, 2025ApJ...985..117L};
see also \citet{2019JCAP...08..011F, 2021arXiv211101150G, 2022JHEP...09..116C}.
\citet{2021JCAP...10..024L} show that
even when the inflationary consistency relation holds ($n_\mathrm{t}=-r/8$),
the stiff-amplified inflationary SGWB
can contribute as large as several percent of the critical density during the RD era,
which offers a novel pathway to alleviating the Hubble tension
by the so-called $H_0-N_\mathrm{eff}$ degeneracy
\citep{2013PhRvD..87h3008H,2019JCAP...10..029S}.

In the following, we briefly review the formalism and the numerical algorithm
for calculating the stiff-amplified inflationary SGWB
in our \verb|stiffGWpy| code.
We define the tensor transfer function as
$T(t,f)\equiv h(t,f)/h_\mathrm{ini}(f)$,
where $h(t,f)$ is the amplitude of the tensor mode of frequency $f$
at cosmic time $t$,
and $h_\mathrm{ini}(f)$ is its initial superhorizon value.
The late-time energy spectrum of the inflationary SGWB is
related to the tensor transfer function by
\begin{equation}\label{eq:Omegagw}
    \Omega_\mathrm{GW}(t,f) \equiv \frac{\ud\Omega_\mathrm{GW}}{\ud\ln f}
    = \Delta_\mathrm{t}^2(f) \frac{(2\pi f)^2\,T^2(t,f)}{12a^2H^2},
\end{equation}
where $a(t)$ is the scale factor and $H(t)=\dot a/a$ is the Hubble parameter.
Throughout the paper, the overdot denotes
the derivative with respect to cosmic time.

Time evolution of the tensor transfer function
is governed by the following wave equation \citep{1974ZhETF..67..825G}:
\begin{equation}\label{eq:transfer}
	\ddot T+3H\,\dot T+({2\pi f}/{a})^2\,T=0,
\end{equation}
where the appearance of the Hubble rate implies that
the expansion history and hence the EoS of the Universe
have a strong impact on the inflationary SGWB.
For modes that reentered the horizon
during an era of a \emph{constant} EoS,
the tensor transfer function follows a simple power law
\citep{2008PhRvD..77f3504B, 2017PhRvD..96f3505L}
and the GW spectrum in the corresponding frequency range can be expressed as
\begin{equation}\label{eq:Omegagw_pl}
    \Omega_\mathrm{GW}(f) \propto 
    \Delta_\mathrm{t}^2(f)\left(\frac{f\,a_f}{H_0}\right)^2
    \propto f^{\,n_\mathrm{t}+\frac{2(3w_f-1)}{1+3w_f}},
\end{equation}
where $a_f$ is the scale factor at which the tensor mode
of frequency $f$ reentered the horizon, $2\pi f\equiv a_f H(a_f)$,
and $w_f$ is the EoS parameter then \citep{2025ApJ...985..117L}.

However, the tensor wave equation~(\ref{eq:transfer})
does not generally have analytical solutions.
Its exact solutions must be obtained by numerical methods.
For this purpose, \citet{2021JCAP...10..024L}
implemented a dynamical system approach,
defining the following dimensionless dynamical variables for each frequency:
\begin{equation}\label{eq:dyn_var}
    \zeta_f \equiv \ln \frac{2\pi f}{aH},\quad x_f \equiv \frac{\dot{T}}{H},
    \quad y_f \equiv \frac{2\pi f}{aH}\, T. 
\end{equation}
Apparently, $T(t,f)=y_f/e^{\zeta_f}$.
Using these variables, one can then rearrange Eq.~(\ref{eq:transfer})
into the following dynamical system:
\begin{IEEEeqnarray}{rCl}
    \zeta_f' & = & \frac{3}{2}\sigma - 1, \label{eq:zeta}\\
    x_f' & = & -3x_f + \frac{3}{2}\sigma \, x_f - e^{\zeta_f}y_f, \label{eq:x}\\
    y_f' & = & -y_f + \frac{3}{2}\sigma \, y_f + e^{\zeta_f} x_f, \label{eq:y}
\end{IEEEeqnarray}
where the prime denotes the derivative with respect to the number of $e$-folds,
$N\equiv \ln a$ ($\ud N = H\,\ud t$), and
\begin{equation}\label{eq:sigma}
    \sigma \equiv -\frac{2\dot H}{3H^2} = \frac{\bar\rho+\bar P}{\bar\rho} = 1+w.
\end{equation}
Since the inflationary SGWB may contribute appreciably
to the total EoS parameter $w$,
its backreaction on the expansion history
is then encoded in the $\sigma$ variable above
through the density fraction of the SGWB,
$\Omega_\mathrm{GW}(N)\equiv\rho_\mathrm{GW}/\bar\rho
=\int \Omega_\mathrm{GW}(N,f)\,\ud\ln f$.
Therefore, Eqs.~(\ref{eq:zeta}--\ref{eq:sigma})
form a set of coupled integro-differential equations.
These equations are solved by the \verb|stiffGWpy| code
for a range of adaptively chosen frequencies.

The cosmological model implemented in \verb|stiffGWpy|
is a simple extension to the base $\Lambda$CDM model.
The thermal history of the Universe can begin with kination
and then transition to RD prior to Big Bang nucleosynthesis (BBN).
Kination is modeled by a stiff fluid ($w_\mathrm{s}=1$) 
and the kination-to-radiation transition is parameterized by 
$\kappa_{10}\equiv (\rho_\mathrm{s}/\rho_\gamma)|_{T=10~\mathrm{MeV}}$, 
the ratio of the stiff-fluid density to the photon density at $10$~MeV
\citep{2010PhRvD..82h3501D}.
The \verb|stiffGWpy| model also assumes that
the inflationary phase can end into a prolonged reheating epoch 
dominated by the coherent oscillations of the inflaton field
so that $w_\mathrm{re}=0$.
The end of reheating (at temperature $T_\mathrm{re}$)
marks the beginning of the thermal history.
In summary, our physical model contains the following free parameters
apart from the usual $\Lambda$CDM parameters:
$\{r, n_\mathrm{t}, \kappa_{10}, T_\mathrm{re}, \Delta N_\mathrm{re}\}$,
where 
$\Delta N_\mathrm{re}$ is the number of $e$-folds during reheating.
Further details of the physical model can be found in Appendix~\ref{app:model}.

\begin{figure}[ht!]
    \epsscale{1.2}
    \plotone{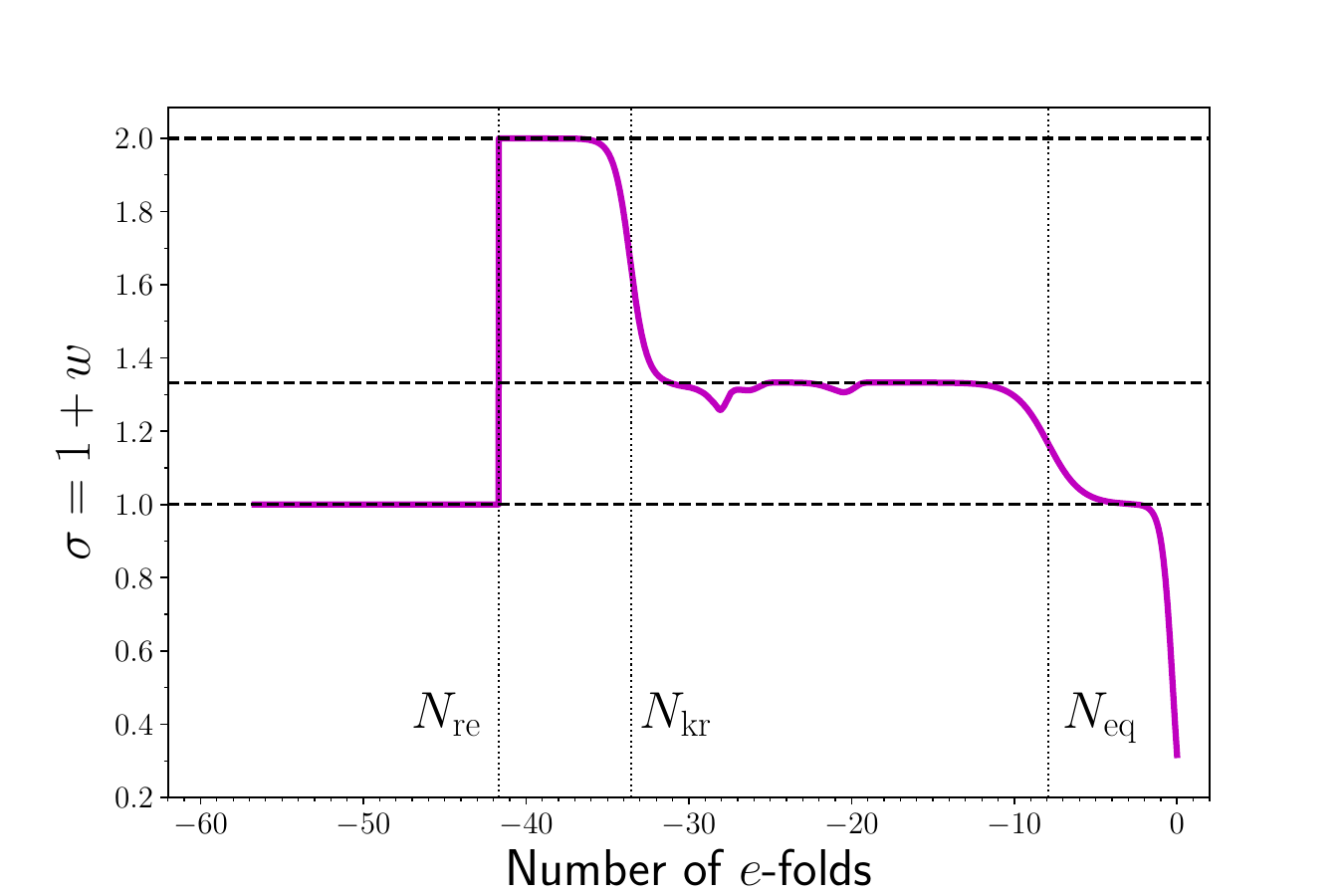}
    \caption{Example evolution of the $\sigma$ variable
    or, equivalently, the EoS parameter.
    The time variable is given by the number of $e$-folds, $N$,
    a measure of the exponential expansion of the Universe.
    The evolution of $\sigma(N)$ signifies the expansion history.
    The horizontal dashed lines mark the EoS parameters
    of the major cosmological phases:
    $w=0$ during the matter-dominated era and reheating,
    $w=1/3$ during the radiation-dominated era, and $w=1$ during kination.
    The features in the $\sigma$ curve during the RD era
    indicate the phase transitions in the thermal history
    of the SM sector; see Appendix~\ref{app:thermal}.
    Key cosmological transitions are marked by the vertical dotted lines:
    $N_\mathrm{re}$ denotes the end of reheating,
    when all the inflationary energy is transformed
    to the energy of the stiff matter and the thermal bath; 
    $N_\mathrm{kr}$ denotes the kination-radiation equality,
    when the energy density of the stiff matter ($\rho_\mathrm{s} \propto a^{-6}$)
    equals that of radiation ($\rho_\mathrm{r} \propto a^{-4}$);
    $N_\mathrm{eq}$ denotes the ordinary radiation-matter equality.
    }\label{fig:sigma}
\end{figure}

\begin{figure*}[ht!]
    \resizebox{\textwidth}{!}{\includegraphics{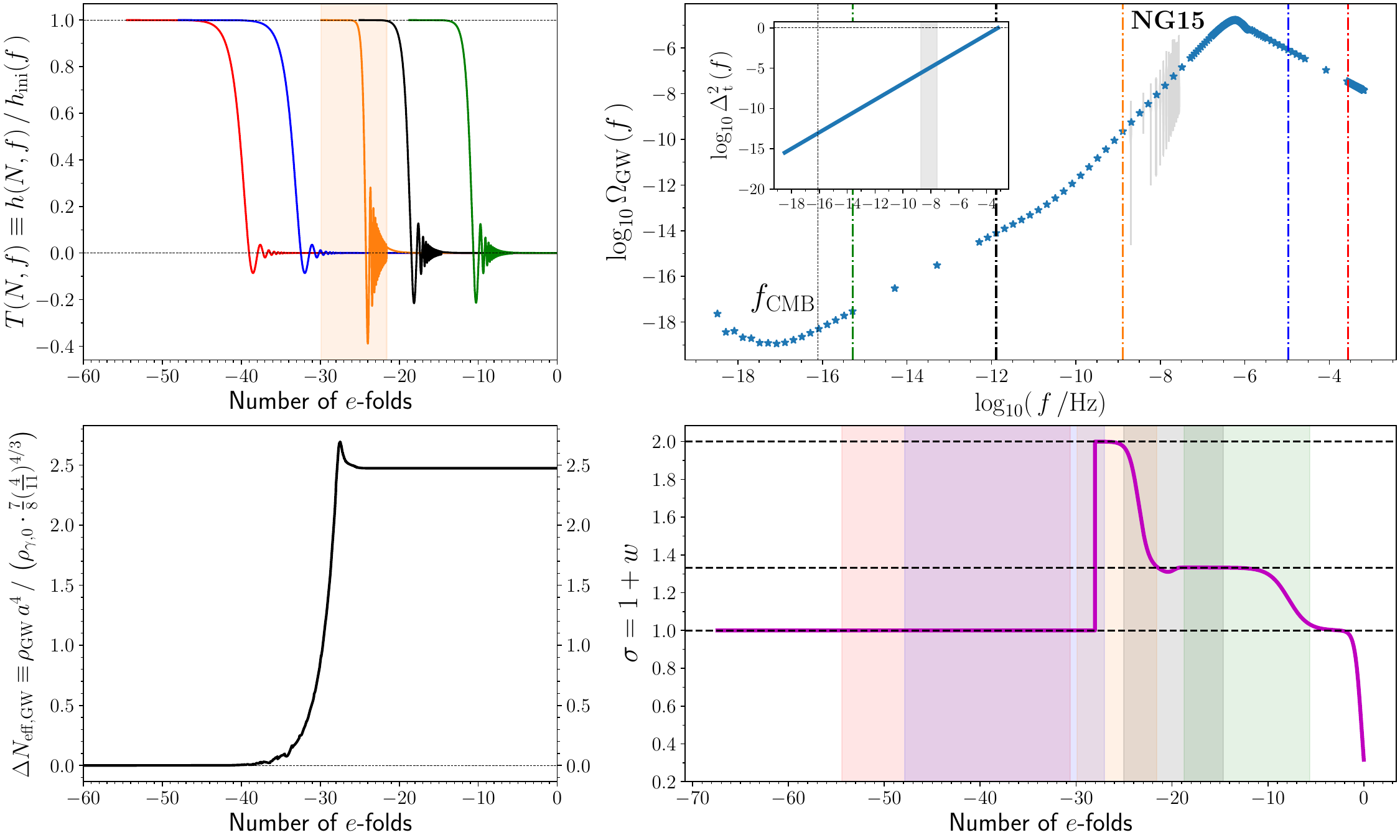}}
    \caption{\emph{Upper left}: tensor transfer functions
    for five illustrative tensor modes.
    The two leftmost modes reentered the horizon during reheating ($w=0$)
    and the two rightmost modes during the RD era ($w=1/3$).
    The middle one reentered the horizon during kination ($w=1$)
    and the vertical shade indicates its interval of integration.
    \emph{Upper right}: Stiff-amplified inflationary SGWB spectrum today.
    As explained in the text, the sampled frequencies, $\{f_i\}$,
    are adaptively chosen such that the sampling is denser
    around transitions in $\Omega_\mathrm{GW}(f)$.
    The gray violins indicate the free spectrum obtained
    from the NANOGrav 15 yr data set \citep[``NG15'',][]{2023ApJ...951L...8A}.
    The vertical dash-dotted lines denote
    the same illustrative frequencies as in the upper left panel.
    The inset shows the blue-tilted primordial tensor spectra,
    where the nonlinearity cutoff is illustrated (cf. Appendix~\ref{app:cutoff}),
    and the vertical shade denotes the PTA frequency band.
    The vertical dotted lines in both this panel and the inset denote the CMB scale, $f_\mathrm{CMB}$.
    \emph{Lower left}: evolution of the effective number of extra relativistic species
    due to the stiff-amplified inflationary SGWB, $\Delta N_\mathrm{eff,GW}$.
    \emph{Lower right}: expansion history described by
    the evolution of the $\sigma$ variable, similar to Fig.~\ref{fig:sigma}.
    The vertical shades indicate the integration time spans of the five illustrative modes.
    }\label{fig:solution}
\end{figure*}

Fig.~\ref{fig:sigma} shows an example of the expansion history
in our \verb|stiffGWpy| model, in terms of the $\sigma$ variable.
The evolution of $\sigma(N)$ exhibits
the important cosmological transitions mentioned above.
Such nontrivial evolutions support our approach
of seeking exact solutions to the tensor wave equation
or the dynamical system~(\ref{eq:zeta}--\ref{eq:y})
for accurate predictions of the inflationary SGWB,
as opposed to using fitting formulae.
Nevertheless, these numerical solutions can be computationally expensive, 
especially because of the large number of modes needed
to capture the accurate shape of $\Omega_\mathrm{GW}(f)$
across the entire frequency range, as illustrated in Fig.~\ref{fig:solution}.
It also illustrates the adaptive frequency sampling strategy implemented in
\verb|stiffGWpy|, i.e., denser samples
around transitions of spectral indices in $\Omega_\mathrm{GW}(f)$.
As a result, an iterative run for a fixed set of model parameters
may take $\mathcal{O}(10)$ seconds in practice.
Given the slow and complex nature of numerical solutions,
developing a fast solver that enables
rapid characterization of the inflationary SGWB
and hence efficient parameter sweep
is essential for analyzing large GW and cosmological data sets.

\section{Design of Emulator}\label{sec:design}

In this section, we first provide arguments for designing
a \emph{neural network} emulator to accelerate the computation
of the stiff-amplified inflationary SGWB,
and then determine the appropriate targets for our deep learning model.

Solving the coupled integro-differential equations~(\ref{eq:zeta}--\ref{eq:sigma})
requires an iterative algorithm in which each iteration involves
numerical integration of a family of ODEs. 
Existing ODE solvers typically rely on computational libraries
such as \verb|SciPy| \citep{scipy},
which utilizes the \verb|LSODA| method \citep{lsoda}. 
This approach is time-intensive 
and lacks support for hardware acceleration, e.g., those for GPUs.
Using GPU acceleration libraries like \verb|CuPy| \citep{cupy_learningsys2017}
can optimize over variable data structures
and achieve some degree of acceleration.
Nonetheless, each iteration of numerical integration
still requires hundreds of time steps (as in the \verb|stiffGWpy| code),
resulting in a lengthy computation process.
To overcome this issue, we propose \verb|SageNet|,
a deep learning-based emulator for \verb|stiffGWpy|
that leverages the nonlinear properties
of neural networks to capture the characteristics of the target curves,
attaining high-fidelity, accelerated predictions
for the inflationary SGWB with possible stiff amplifcation.
This is arguably the most feasible approach.

The nonlinear, end-to-end capabilities of neural networks
offer two possible methods to design the emulator.
The first method is to fit the solution
to the dynamical system (\ref{eq:zeta}--\ref{eq:y})
for a given expansion history, $\sigma(N)$, for each frequency.
As described in Appendix~\ref{app:numerical}, we integrate the ODE system
for a fixed interval of $[\zeta_\mathrm{min}, \zeta_\mathrm{max}]$ for all frequencies.
The resultant solutions are the fitting target in this case,
from which we can then extract the subhorizon amplitude
of the GW spectrum at the end of the integration time span
by the following relation:
\begin{equation}\label{eq:omegaGW}
    \Omega_\mathrm{GW}(N,f)
    = \frac{\Delta^2_\mathrm{t}(f)}{24}\left[x_f^2(N)+y_f^2(N)\right].
\end{equation}
This approach mirrors the traditional numerical methods
used in the \verb|stiffGWpy| code,
which provides the flexibility of deriving evolutionary paths
for various physical quantities via $(\zeta_f,x_f,y_f)$.
However, 
it still requires solving the full integro-differential system iteratively
to obtain the self-consistent result of $\Omega_\mathrm{GW}(N,f_i)$
for a range of frequencies $\{f_i\}$.
This increases computational complexity substantially.

The second method directly sets the present-day SGWB spectrum,
$\Omega_\mathrm{GW}(f)$, as the learning target. 
Even though the sampled frequencies $\{f_i\}$
and their limits $(f_\mathrm{min},f_\mathrm{max})$
are adaptively chosen according to the parameters of the physical model
(see the upper right panel of Fig.~\ref{fig:solution}),
neural networks are able to reconstruct the variable distribution of $\{f_i\}$
and, at the same time, $\{\Omega_\mathrm{GW}(f_i)\}$.
This curve typically consists of only hundreds of points.
As a result, the learning task for neural networks is simpler
than in the first method above.
Potential computational complexity can arise from
the significant fluctuations of $f_\mathrm{max}$
due to the UV cutoff in the primordial tensor spectrum; see Appendix~\ref{app:cutoff}.
Nevertheless, by circumventing intermediate calculations,
this approach reduces the overall computational cost considerably
compared with the first method that fits the evolutions of $x_f$ and $y_f$.
Therefore, we adopt the second method
to design the neural network emulator for the inflationary SGWB in this work.

\section{Networks and Training}\label{sec:network}

In this section, we first explain how we generate the data set,
and then describe the implementation
of the LSTM architecture
in our neural network emulator, \verb|SageNet|.
The training process is reported at the end of the section.

\subsection{Data Preparation}

\textbf{Parameter Sampling.} We generate the data set
based on the physical model of the stiff-amplified inflationary SGWB
described in Section~\ref{sec:SGWB}.
In particular, we use the \verb|stiffGWpy| code to calculate the GW spectra,
$\{f_i,\,\log_{10}\Omega_\mathrm{GW}(f_i)\}$, 
for different sets of model parameters iteratively. 
These samples are defined by five physical parameters,
listed in Table \ref{table:parameter_range},
each following a uniform distribution within its specified range.

\begin{table}[htbp]
    \centering
    \caption{Prior ranges of the sampled parameters
    of the stiff-amplified inflationary SGWB model. 
    We adopt uniform distributions for all parameters;
    cf. \citet{2025ApJ...985..117L}.
    }\label{table:parameter_range}
    \small
    \setlength{\tabcolsep}{2.5pt}
    \renewcommand{\arraystretch}{1.1}
    \begin{tabular}{cccccc}
        \toprule
        \textbf{Param}&\textbf{$\log_{10}r$} & \textbf{$n_\mathrm{t}$} 
        & \textbf{$\log _{10}\kappa _{10}$} 
        & \textbf{$\log _{10}( T_{\mathrm{re}}/\mathrm{GeV})$} & \textbf{$\Delta N_{{\mathrm{re}}}$}\\
        \midrule
        \textbf{Limits}& $[-25,0]$      & $[-1,6]$ & $[-7,3]$ & $[-3,7]$ & $[0,40]$ \\
        \textbf{Prior}&&& Uniform && \\
        \bottomrule
    \end{tabular}
\end{table}

\begin{figure}[ht!]
    \plotone{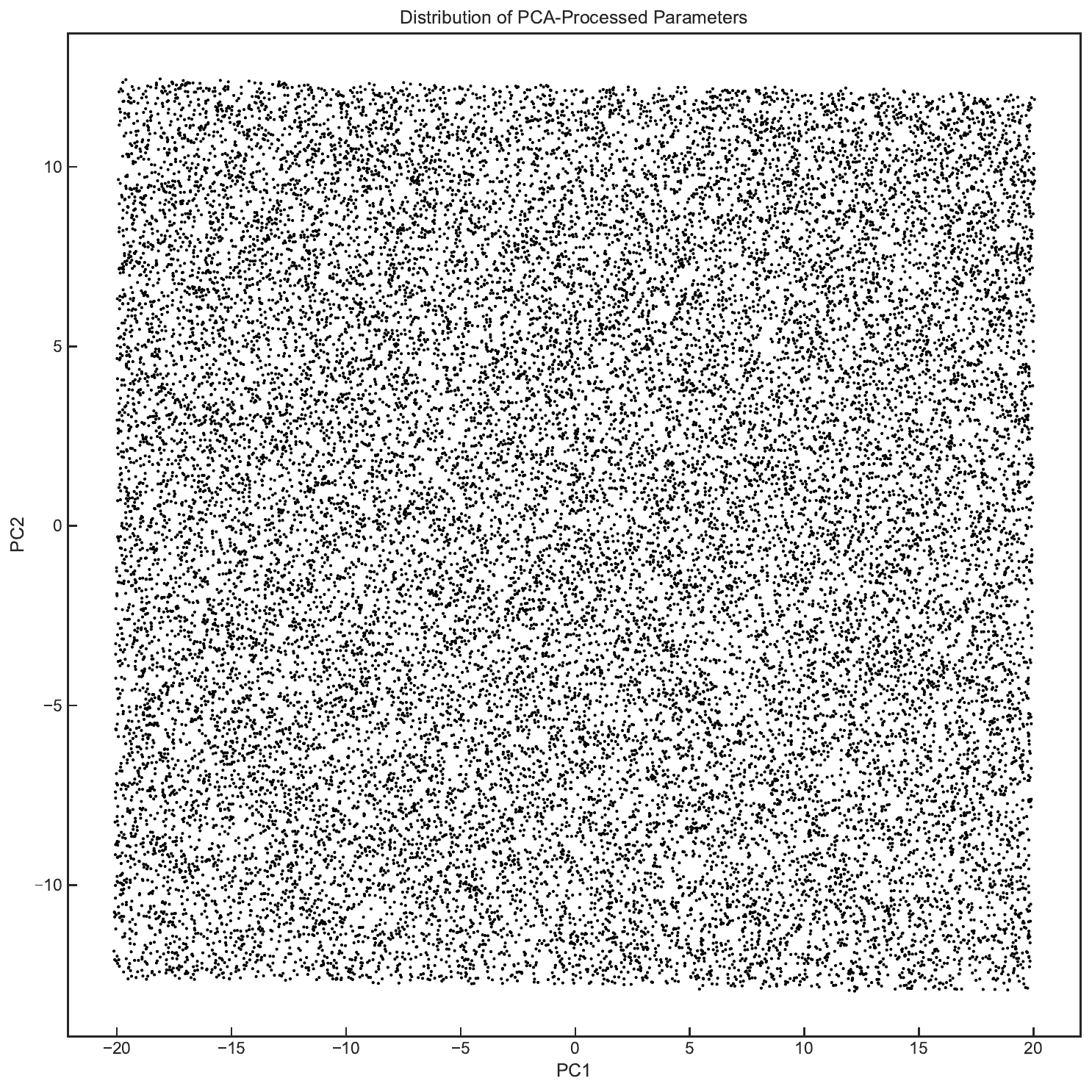}
    \caption{Projected distribution of all samples in the data set
    after applying Principal Component Analysis (PCA).
    We perform a dense search of the parameter space
    of $\{r, n_\mathrm{t}, \kappa_{10}, T_\mathrm{re}, \Delta N_\mathrm{re}\}$
    using a data set of 25,689 samples generated by LHS
    with their priors specified in Table~\ref{table:parameter_range}.
    The resultant distribution after PCA processing
    is visualized by projecting the five-dimensional space
    onto a two-dimensional plane defined by the first two principal components.}
    \label{fig:parameter distribution}
\end{figure}

We employ Latin Hypercube Sampling (LHS)
to achieve uniform coverage of the five-dimensional parameter space
with enhanced space filling and convergence efficiency.
Approximately 25,000 samples of the SGWB parameters are generated,
and their distributions are visualized in Fig.~\ref{fig:parameter distribution}.
It exhibits a uniform distribution of the projected samples,
thus supporting the validity of our sampling approach.

\begin{figure}[ht!]
    \centering
    \includegraphics[width=1\linewidth]{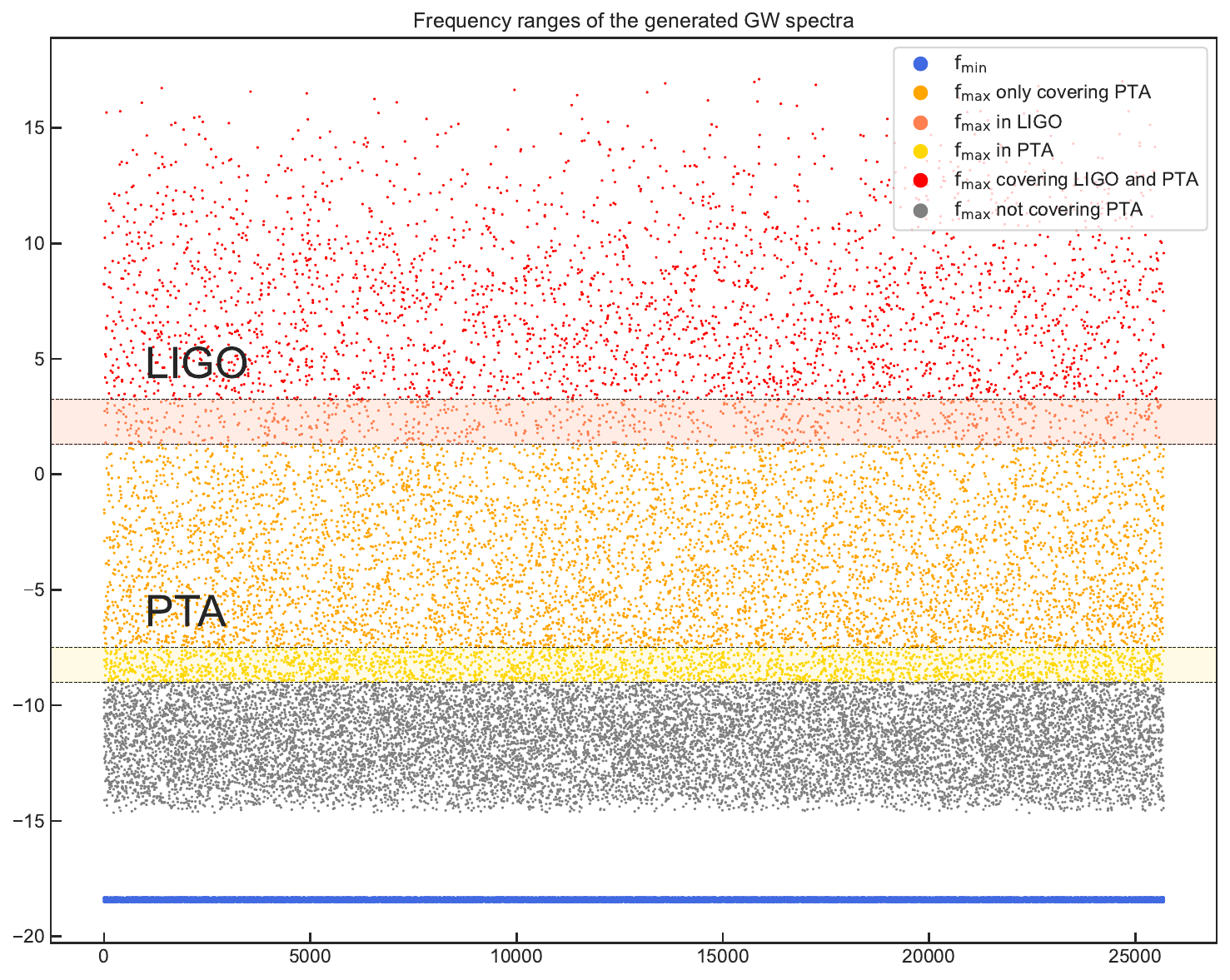}
    \caption{Frequency ranges of $\{f_i\}$,
    or $[f_\mathrm{min},f_\mathrm{max}]$, of all the samples in our data set.
    They vary inherently across the samples
    and their distributions are statistically analyzed here.
    The horizontal axis represents sample indices,
    and the vertical axis indicates the frequency boundaries derived from each set of physical parameter set.
    We find that $f_\mathrm{min}$ remains stable near $10^{-18.6}$\,Hz, forming an approximate horizontal line,
    while $f_\mathrm{max}$ exceeds $10^{-15}$\,Hz for all samples, distinguishable from the former.}
    \label{fig:frequency range}
\end{figure}

\begin{figure}[ht!]
    \centering
    \includegraphics[width=1\linewidth]{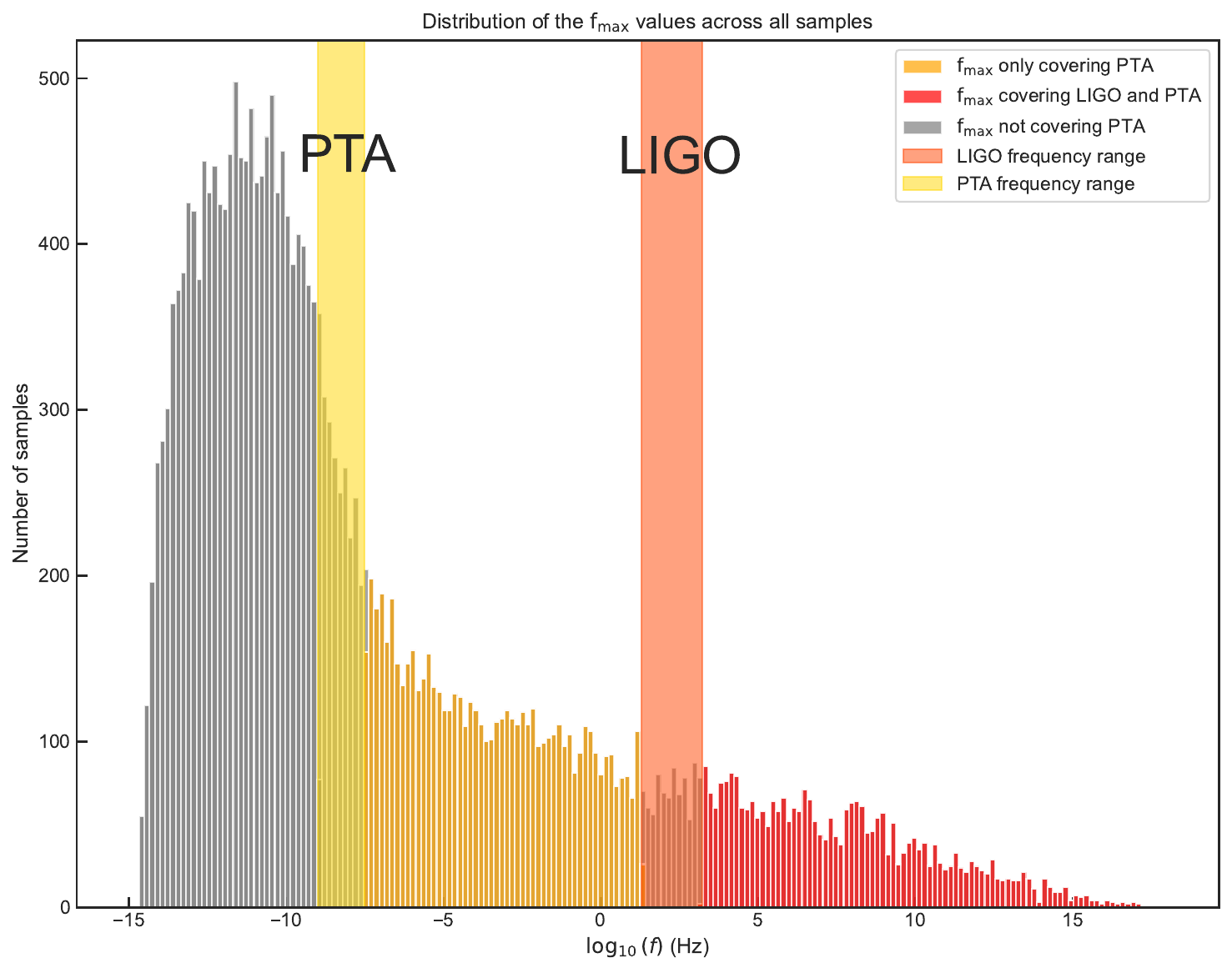}
    \caption{Histogram of the $f_\mathrm{max}$ values of all samples in the data set. 
    Same as in Fig. \ref{fig:frequency range}, the orange and red regions represent the samples
    that cover the PTA frequency range and the LIGO frequency range, respectively.}
    \label{fig:frequency range hist}
\end{figure}

\textbf{Data Generation.} Using multithreading,
we iteratively compute the frequencies $\{f_i\}$
and the $\{\log_{10}\Omega_\mathrm{GW}(f_i)\}$ curve for each sample.
This process is executed on an Intel Xeon Platinum 8352V processor,
which take approximately 6 hours to generate all the 25,000 samples.
As illustrated in Fig.~\ref{fig:frequency range}, 
60.74\% of the spectra, $\Omega_\mathrm{GW}(f)$,
cannot reach the PTA frequency band ($f_\mathrm{max} < f_\mathrm{PTA,\,{min}}$),
which is natural consequence of the UV cutoff implemented in \verb|stiffGWpy|;
see Appendix~\ref{app:cutoff}.
Meanwhile, 11.56\% of the spectra can cover as far as the LIGO frequency band
($f_\mathrm{max} > f_\mathrm{LIGO,\,{max}}$).
Fig.~\ref{fig:frequency range hist} shows the histogram of the $f_\mathrm{max}$ values
of all samples in the data set.

\begin{figure*}
    \centering
    \includegraphics[width=0.78\linewidth]{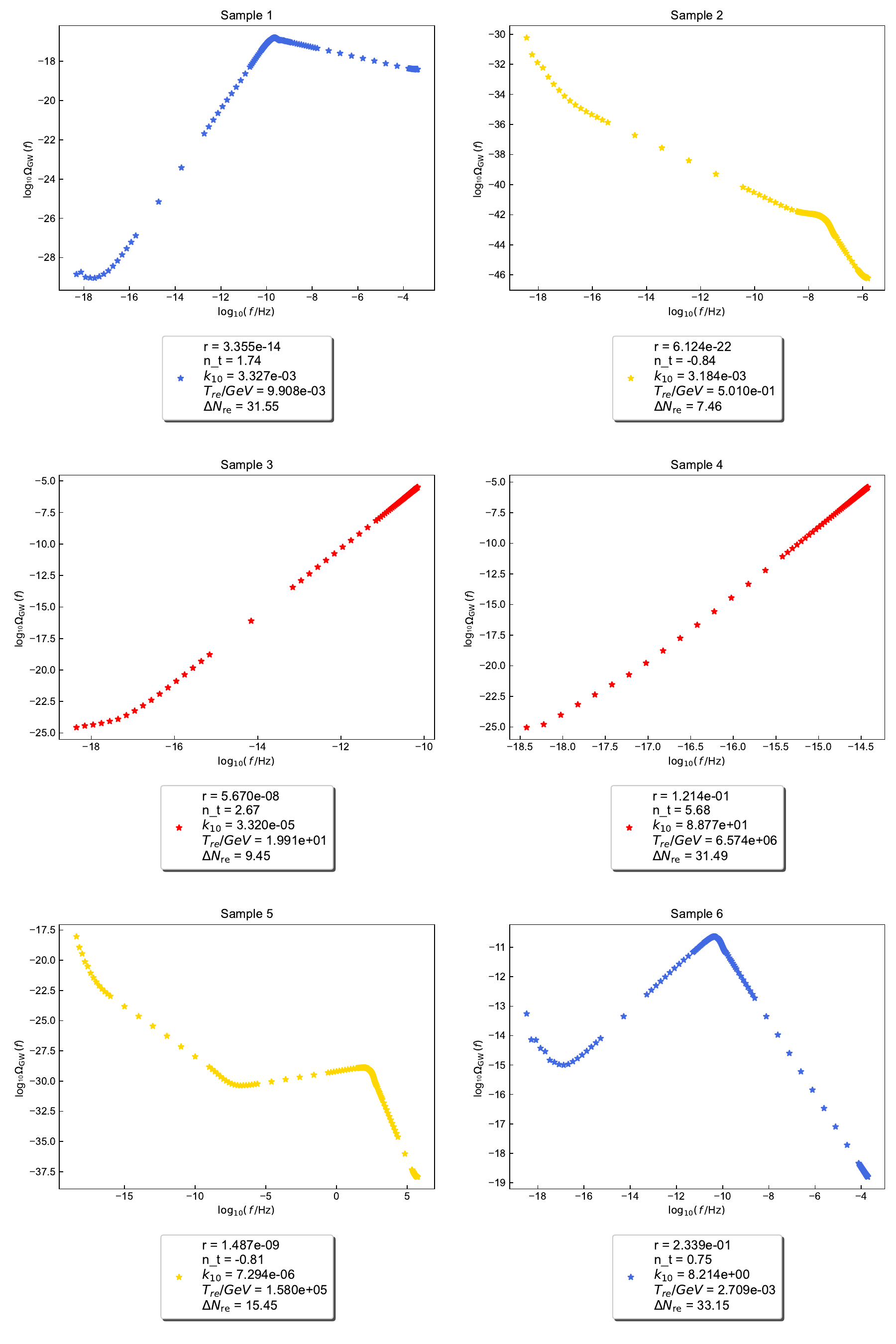}
    \caption{Sampled $\{f_i, \Omega_\mathrm{GW}(f_i)\}$ curves.
    This figure displays six representative $\Omega_\mathrm{GW}(f_i)$ curves, illustrating three distinct morphological types defined by their monotonicity and frequency characteristics. The first type exhibits a rapid monotonic increase followed by an abrupt cutoff. The second type shows a monotonic decrease with a variable inflection point, where the rate of decrease shifts noticeably. The third, most complex type features multiple monotonicity transitions—initially decreasing, then increasing, and finally decreasing again—posing significant challenges for prediction due to variability in monotonicity, frequency range, and $\Omega_\mathrm{GW}(f_i)$ amplitude.
    Different colors distinguish these types:
    red for monotonically increasing curves,
    yellow for monotonically decreasing (or nearly monotonic) curves,
    and blue for curves without monotonicity.}
    \label{fig:sample of omega GW}
\end{figure*}

Representative curves of $\log_{10}\Omega_\mathrm{GW}(f)$
are presented in Fig.~\ref{fig:sample of omega GW}.
Variations in the parameters of the physical model
result in three primary morphological characteristics, described as follows:
\begin{itemize}
\item Monotonically increasing behavior with relatively low cutoff frequencies,
where most curves satisfy $f_\mathrm{max} < f_\mathrm{PTA,\,{min}}$.
\item Non-monotonic behavior, with most curves exhibiting a distinct maximum
at $f_\mathrm{re}$, the frequency of the mode that fills the horizon at $T_\mathrm{re}$.
\item Monotonically decreasing behavior.
\end{itemize}

\begin{figure}[ht!]
    \centering
    \includegraphics[width=1\linewidth]{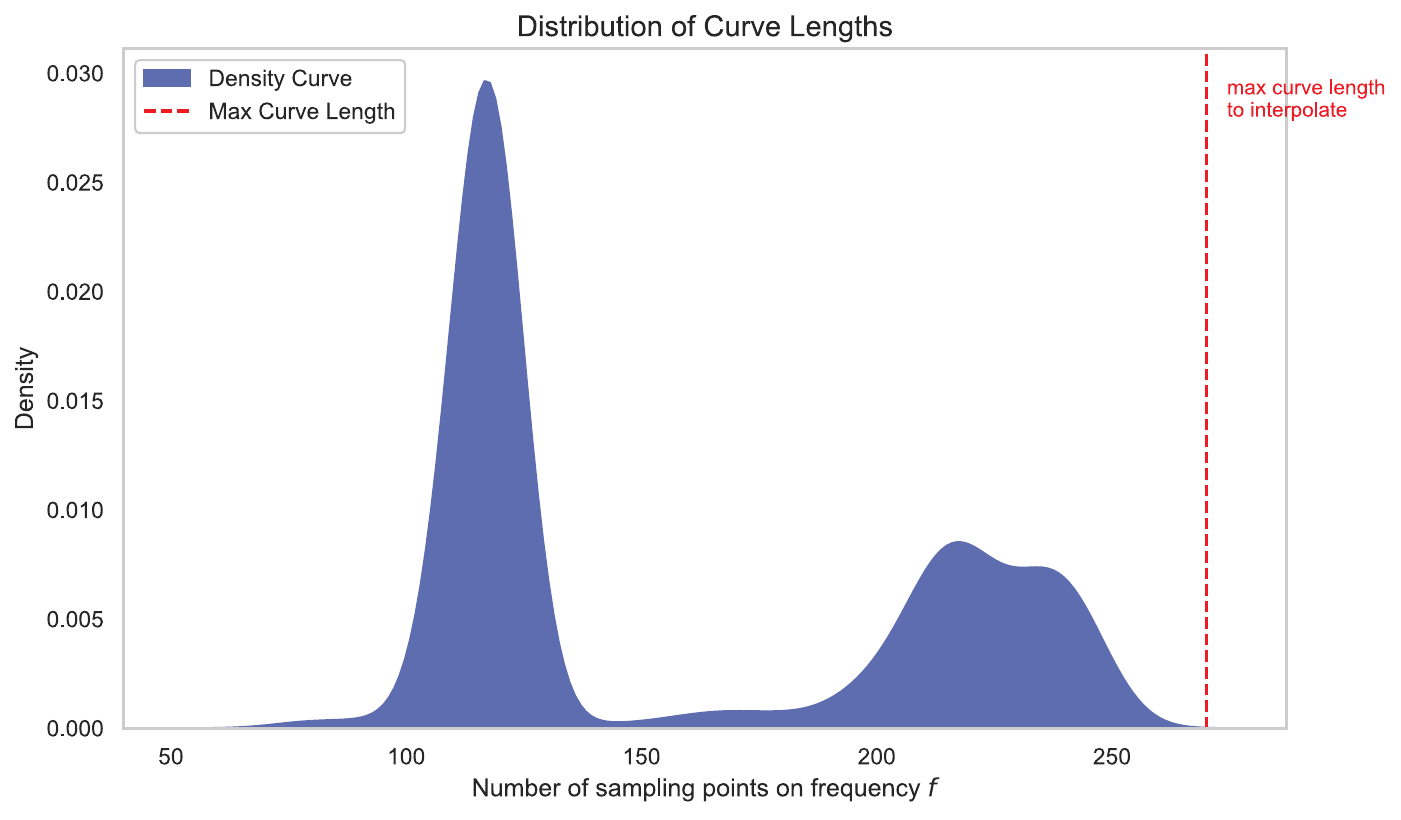}
    \caption{Distribution of the number of sampled points/frequencies
    in all GW spectral curves in the data set, $\{\Omega_\mathrm{GW}(f_i)\}$,
    generated by stiffGWpy.
    }\label{fig:dist_length}
\end{figure}

Meanwhile, the resultant curves of $\{\Omega_\mathrm{GW}(f_i)\}$
have nonuniform lengths (numbers of representative points/frequencies).
This is not surprising since we implement an adaptive sampling strategy
to compute the GW spectrum curves in the \verb|stiffGWpy| code,
as described in Section~\ref{sec:SGWB}.
The distribution of the number of sampled frequencies per curve
is illustrated in Fig.~\ref{fig:dist_length}.
Interestingly, the count of sampled points
mostly falls within two intervals: $[110,125]$ and $[200,256]$.

\textbf{Data Interpolation.} The nonuniform sampling density of $\{f_i\}$
in the theoretical calculation is related to the changes of spectral indices
in the $\Omega_\mathrm{GW}(f)$ curves,
with denser sampling around steep changes; see Fig.~\ref{fig:solution}.
However, variable lengths of the input data curves can impede model learning.
To tackle this problem, we apply univariate interpolation
to insert points between the two closest points
in the raw theoretical curve of $\{\Omega_\mathrm{GW}(f_i)\}$,
so that each curve in the training set has a uniform length of 256 points.

This approach to standardization of the curve length
is helpful because it preserves the sampling distribution
in the raw curves, which contains artificial information
introduced by the adaptive sampling method in the \verb|stiffGWpy| code.
Otherwise, inserting points into wider intervals of $[f_i, f_{i+1}]$
would alter the density distribution of sampled frequencies,
potentially losing this artificial information.
Therefore, by applying the interpolation scheme above,
we preserve both the intrinsic curve characteristics
and the artificially introduced sampling density features,
which allows us to examine whether the emulator
can learn both attributes effectively.

\subsection{Network Architecture}

To address the sequential nature of the GW energy density spectra, $\Omega_{\mathrm{GW}}(f)$,
we design our neural network emulator, \verb|SageNet|,
with an LSTM architecture as its core component.
Before detailing the LSTM structure, we first clarify the distinction
between RNNs and non-RNNs
and explain why recurrent architectures, particularly LSTM, are well-suited for our task.

\textbf{RNN versus non-RNNs.} Neural networks can be broadly categorized
into recurrent and non-recurrent architectures based on their ability to process sequential data.
Non-recurrent networks, such as fully connected deep neural networks (DNNs)
or convolutional neural networks (CNNs),
treat inputs as independent data points without explicit modeling of temporal or sequential dependencies.
For a given input vector $\mathbf{x} \in \mathbb{R}^d$,
a DNN computes the output through a series of feedforward transformations:
\begin{equation}
	\mathbf{y} = \sigma_L(W_L \cdot \sigma_{L-1}(W_{L-1} \cdot \ldots 
	\sigma_1(W_1 \cdot \mathbf{x} + \mathbf{b}_1) \ldots ) + 	\mathbf{b}_L),
\end{equation}
where $W_L$ and $\mathbf{b}_L$ are the weight matrix and the bias vector for layer $L$,
and $\sigma_L(\cdot)$ is the activation function. 
This structure assumes that all input dimensions are equally weighted
and lacks mechanisms to capture relationships between sequential elements,
making it less effective for tasks where the order of data points matters,
such as the frequency-dependent $\Omega_{\mathrm{GW}}(f)$ curves in our study.

In contrast, recurrent neural networks (RNNs)
are designed to process sequences by maintaining a hidden state that evolves with each time step,
effectively encoding temporal dependencies.
For a sequence of inputs $\{\mathbf{x}_1, \mathbf{x}_2, \ldots, \mathbf{x}_T\}$,
an RNN updates its hidden state $\mathbf{h}_t$ at time step $t$ as:
\begin{equation}
\mathbf{h}_t = \sigma(W_h \cdot [\mathbf{h}_{t-1}, \mathbf{x}_t] + \mathbf{b}_h),
\end{equation}
where $\mathbf{h}_{t-1}$ is the previous hidden state,
$W_h$ and $\mathbf{b}_h$ are the weight matrix and the bias,
and $\sigma(\cdot)$ is typically a nonlinear activation (e.g., $\tanh$).
The hidden state $\mathbf{h}_t$ acts as a memory,
allowing the network to incorporate information from previous time steps when processing $\mathbf{x}_t$.
This recursive mechanism is particularly advantageous for our task,
where $\Omega_{\mathrm{GW}}(f)$ is represented as a sequence of frequency points
$\{f_i, \Omega_{\mathrm{GW}}(f_i)\}$ with adaptive sampling (see Fig.~\ref{fig:solution}).
The spectral shape, including transitions in spectral indices due to cosmological phases
(e.g., kination or radiation domination), depends on the relationships between adjacent frequency points.
RNNs can model these relationships by propagating contextual information across the sequence,
enabling accurate reconstruction of the morphology of the curve.
In comparison, non-recurrent architectures, such as DNNs, however,
struggle to capture these sequential dependencies without significant preprocessing,
which may lead to loss of critical information.
CNNs, while capable of extracting local patterns, are less effective
at modeling long-range dependencies across the entire frequency range
that can span as wide as 30 orders of magnitude.
Thus, the recurrent nature of RNNs provides a significant advantage for our application
due to the sequential nature of $\Omega_{\mathrm{GW}}(f)$.

\textbf{LSTM Structure.} As explained above, RNNs have been widely used in forecasting sequential data
and are thus suitable for the type of data in this study.
However, 
standard RNNs suffer from the vanishing or exploding gradient problem during backpropagation,
which hinders their ability to learn long-term dependencies in sequences \citep{hochreiter_long_1997}.
In our case, the ability to model long-term dependencies is critical for accurately predicting spectral features,
such as the steep changes in $\Omega_{\mathrm{GW}}(f)$ due to cosmological transitions.
Meanwhile, standard RNNs often fail to maintain the necessary contextual information across the sequence,
leading to degraded performance.
These shortcomings prompt us to use LSTM instead,
an advanced variant of RNNs proposed by \citet{hochreiter_long_1997},
as the core component of \verb|SageNet|
to ensure robust and accurate emulation of the GW spectra.

\begin{figure}[ht!]
    \centering
    \includegraphics[width=1\linewidth]{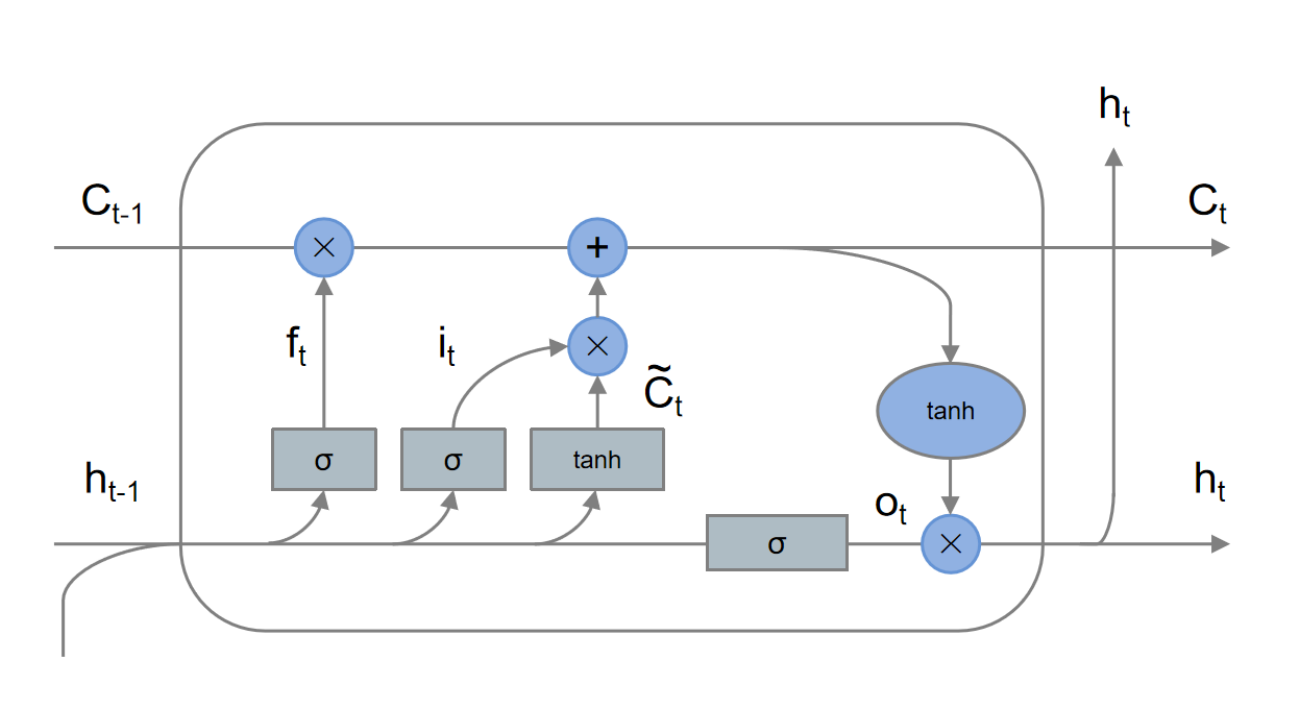}
    \caption{Structure of an LSTM unit.}
    \label{fig:lstm}
\end{figure}

The LSTM network addresses the issue of vanishing or exploding gradients
that degrade traditional RNN performance when processing long sequences.
By incorporating a memory cell (cell state)
and three gating mechanisms--the forget gate, input gate,
and output gate--LSTMs can effectively capture long-term dependencies in time-series data.
Given the sequential nature of the $\{\Omega_\mathrm{GW}(f_i)\}$ curves we aim to predict,
we use an LSTM-based architecture as the primary model in \verb|SageNet|
to learn the characteristics of the curves.
The structure of an LSTM unit is illustrated in Fig.~\ref{fig:lstm}.

The core component of an LSTM is a memory cell regulated by three gating mechanisms:
the forget gate, input gate, and output gate.
These components work together to process the current input $x_t$,
the previous hidden state $h_{t-1}$, and the previous cell state $C_{t-1}$,
controlling the retention, updating, and output of information.
The mathematical formulations are detailed below.
The forget gate determines which information from the previous cell state $C_{t-1}$ to discard.
It is computed as
\begin{equation}
    f_t=\sigma\,(W_f\cdot[h_{t-1},x_t]+b_f).
\end{equation}
Here, $\sigma$ is the sigmoid activation function
(not to be confused with the physical $\sigma$ variable defined in Eq.~[\ref{eq:sigma}]),
producing outputs in the range $[0,1]$; $[h_{t-1}, x_t]$
denotes the concatenation of $h_{t-1}$ and $x_t$; 
and $W_f$ and $b_f$ are the weight matrix and bias vector of the forget gate, respectively.
Values of $f_t$ near 1 indicate retention,
while values near 0 indicate discarding.

The input gate controls how much new information is added to the cell state,
calculated in two steps. The first step is to activate the input gate:
\begin{equation}
    i_t=\sigma(W_i\cdot[h_{t-1},x_t]+b_i),
\end{equation}
and the second step is to generate the candidate cell state:
\begin{equation}
    \tilde{C}_t=\tanh(W_C\cdot[h_{t-1},x_t]+b_C).
\end{equation}
Here, $i_t$, ranging from $[0,1]$, is the output of the input gate;
$\tilde{C}_t$, ranging from $[-1,1]$,
is the candidate update generated by the $\tanh$ activation function;
and $W_i$, $b_i$, $W_C$, and $b_C$ represent the weights and biases
for the input gate and candidate cell state, respectively.
The cell state $C_t$ is updated as:
\begin{equation}
    C_t=f_t\cdot C_{t-1}+i_t\cdot\tilde{C}_t.
\end{equation}
This equation highlights a key feature of LSTM:
$f_t \cdot C_{t-1}$ retains portions of the previous memory,
while $i_t \cdot \tilde{C}_t$ incorporates new information.
The additive structure allows $C_t$ to propagate long-term information
linearly over time steps.

The output gate determines the current hidden state $h_t$.
It is computed as
\begin{IEEEeqnarray}{rCl}
    o_t & = & \sigma(W_o\cdot[h_{t-1},x_t]+b_o),\\
    h_t & = &o_t\cdot\tanh(C_t).
\end{IEEEeqnarray}
Here, $o_t$, ranging from $[0,1]$,
controls the proportion of information output from $C_t$,
while $\tanh(C_t)$ maps the cell state to $[-1,1]$.
The hidden state $h_t$ serves as both the current output
and the input to the next time step.

\begin{figure}[ht!]
    \centering
    \includegraphics[width=1\linewidth]{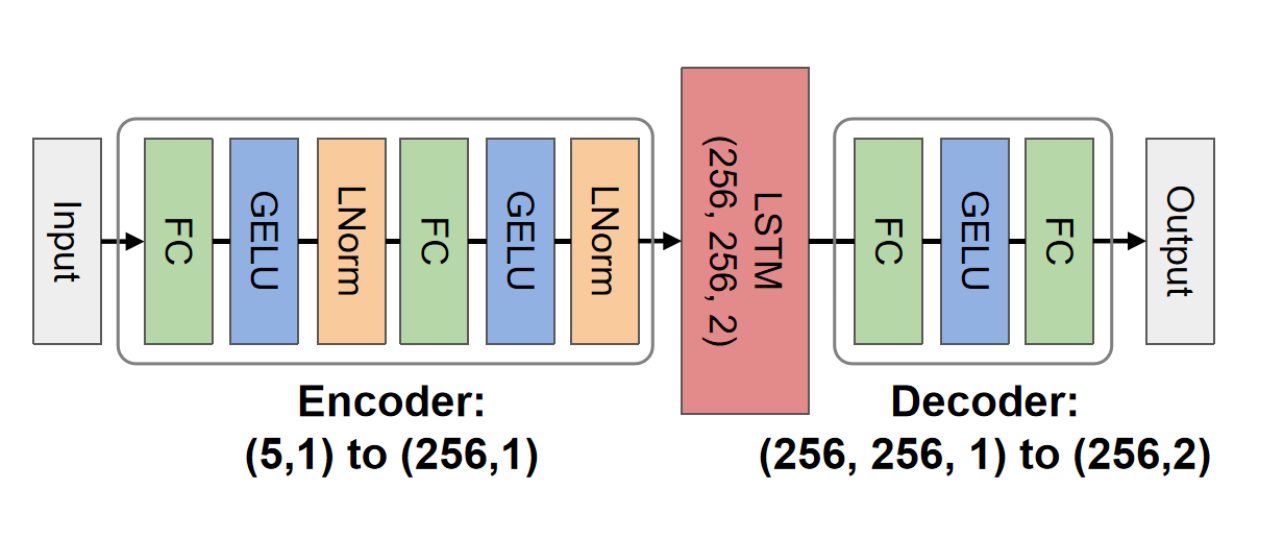}
    \caption{Architecture of SageNet.}
    \label{fig:network}
\end{figure}

\textbf{Our Network.} In \verb|SageNet|, we utilize the network structure
outlined in Table~\ref{tab:model structure} for fitting.
Its architecture is illustrated in Fig.~\ref{fig:network}.
The architecture begins with a parameter encoder
that processes the five-dimensional input parameters,
$\{r, n_\mathrm{t}, \kappa_{10}, T_\mathrm{re}, \Delta N_\mathrm{re}\}$,
using a series of dense layers with Gaussian Error Linear Unit (GELU) activation
and layer normalization:
\begin{IEEEeqnarray}{rCl}
    f_{\mathrm{encode}_j}(x) & = & \mathrm{LayerNorm}(\mathrm{GELU}(\mathrm{W}_{j}\cdot\mathbf{x}+\mathbf{b}_{j})),\quad\\
    \mathbf{h}_{0} & = & f_\mathrm{encode_{2}}(f_\mathrm{encode_{1}}(x)),
\end{IEEEeqnarray}
where $\mathbf{x} \in \mathbb{R}^5$ represents the input parameters, $j=1,2$,
$\mathrm{W}_1 \in \mathbb{R}^{128 \times 5}$ and $\mathrm{W}_2 \in \mathbb{R}^{256 \times 128}$ are weight matrices,
and $\mathbf{b}_1$ and $\mathbf{b}_2$ are their corresponding bias terms.
The encoded representation $\mathbf{h}_0 \in \mathbb{R}^{256}$
is then expanded into a sequence ${h_0, h_0, \dots, h_0} \in \mathbb{R}^{256 \times 256}$
through dimensional repetition to initialize LSTM processing.
The LSTM module processes this sequence via consecutive cell states $c_t$
and hidden states $h_t$ using the gating mechanisms described earlier.
Subsequently, a decoder network projects the LSTM outputs into the target space,
where $\mathrm{W}_3 \in \mathbb{R}^{128 \times 256}$
and $\mathrm{W}4 \in \mathbb{R}^{2 \times 128}$
transform the hidden states into the predicted pairs,
$\{f_i, \Omega_\mathrm{GW}(f_i)\}$:
\begin{equation}
    {\mathbf{\hat{y}_t}}=\mathrm{W}_4\cdot\mathrm{GELU}(\mathrm{W}_3\cdot\mathbf{h_t}+\mathbf{b}_3)+\mathbf{b}_4.
\end{equation}

\begin{table}[ht]
\centering
\caption{Structure of SageNet.}
\label{tab:model structure}
\small
\setlength{\tabcolsep}{1.5pt}
\renewcommand{\arraystretch}{1.1}
\begin{tabular}{lccc}
\toprule
\textbf{Layer} & \textbf{Input Shape} & \textbf{Output Shape} & \textbf{Parameters} \\ 
\midrule
$\textbf{Encoder}$&&&\\
Linear     & [\textit{Batch}, 5] & [\textit{Batch}, 128]             & 896        \\
GELU       & [\textit{Batch}, 128] & [\textit{Batch}, 128]             & 0          \\
LayerNorm  & [\textit{Batch}, 128] & [\textit{Batch}, 128]             & 256        \\
Linear     & [\textit{Batch}, 128] & [\textit{Batch}, 256]             & 33,024     \\
GELU       & [\textit{Batch}, 256] & [\textit{Batch}, 256]             & 0          \\
LayerNorm  & [\textit{Batch}, 256] & [\textit{Batch}, 256]             & 512        \\
$\textbf{Network}$&&&\\
LSTM       & [\textit{Batch}, 256, 256] & [\textit{Batch}, 256, 256]        & 1,048,576  \\
$\textbf{Decoder}$&&&\\
Linear     & [\textit{Batch}, 256, 256] & [\textit{Batch}, 256, 128]        & 32,896     \\
GELU       & [\textit{Batch}, 256, 128] & [\textit{Batch}, 256, 128]        & 0          \\
Linear     & [\textit{Batch}, 256, 128] & [\textit{Batch}, 256, 2]          & 258        \\
\midrule
\textbf{Total}        &          &              & 1,116,418  \\
\bottomrule
\end{tabular}
\end{table}

\subsection{Training}
The emulator is trained using the Mean Squared Error (MSE) loss function
with a curriculum learning approach,
prioritizing accurate prediction of the endpoints of the GW spectra
through weighted loss components.
During inference, \verb|SageNet| applies inverse scaling transformations
to recover physical quantities from the normalized predictions.
Since the curves are generated via iterative computations over a range of frequencies,
we employ an LSTM structure to capture frequency-dependent relationships,
with sequential outputs derived from the hidden states of LSTM.

Before being input into the network, both the parameters and the curves
are standardized by subtracting the mean and scaling to unit variance:
\begin{equation}
	x_{\mathrm{scaled}}=\frac{x-\mu_x}{\sigma_x},\quad x\in\{p,\,c_f,\,c_{\log_{10}\Omega_\mathrm{GW}}\}.
\end{equation}
Each set of the physical parameters of the stiff-amplified inflationary SGWB, $p \in \mathbb{R}^5$,
is mapped to a 256-dimensional feature space,
The process is then repeated 256 times to form a $[256, 256]$ tensor as input to the LSTM.
After processing through two LSTM layers,
the outputs are decoded into a two-dimensional space,
representing the predicted values
of $f$ and $\log_{10}\Omega_{\mathrm{GW}}(f)$ at each time step.
These predictions are compared with the target curves $c \in \mathbb{R}^{256 \times 2}$ using the following loss function:
\begin{equation}
	\mathrm{Loss} = \frac{1}{256B}\sum_{b=1}^B\sum_{i=1}^{256}
	\left[\mathrm{MSE}_f+\mathrm{MSE}_{\log_{10}\Omega_\mathrm{GW}}\right],
\end{equation}
where $B$ is the batch size, $\mathrm{MSE}_f = (\hat{f}_{bi}-f_{bi})^2$,
and $\mathrm{MSE}_{\log_{10}\Omega_\mathrm{GW}}
=({\log_{10}\hat\Omega_\mathrm{GW}}_{bi}-{\log_{10}\Omega_\mathrm{GW}}_{bi})^2$.

\begin{figure}[ht!]
    \centering
    \includegraphics[width=1\linewidth]{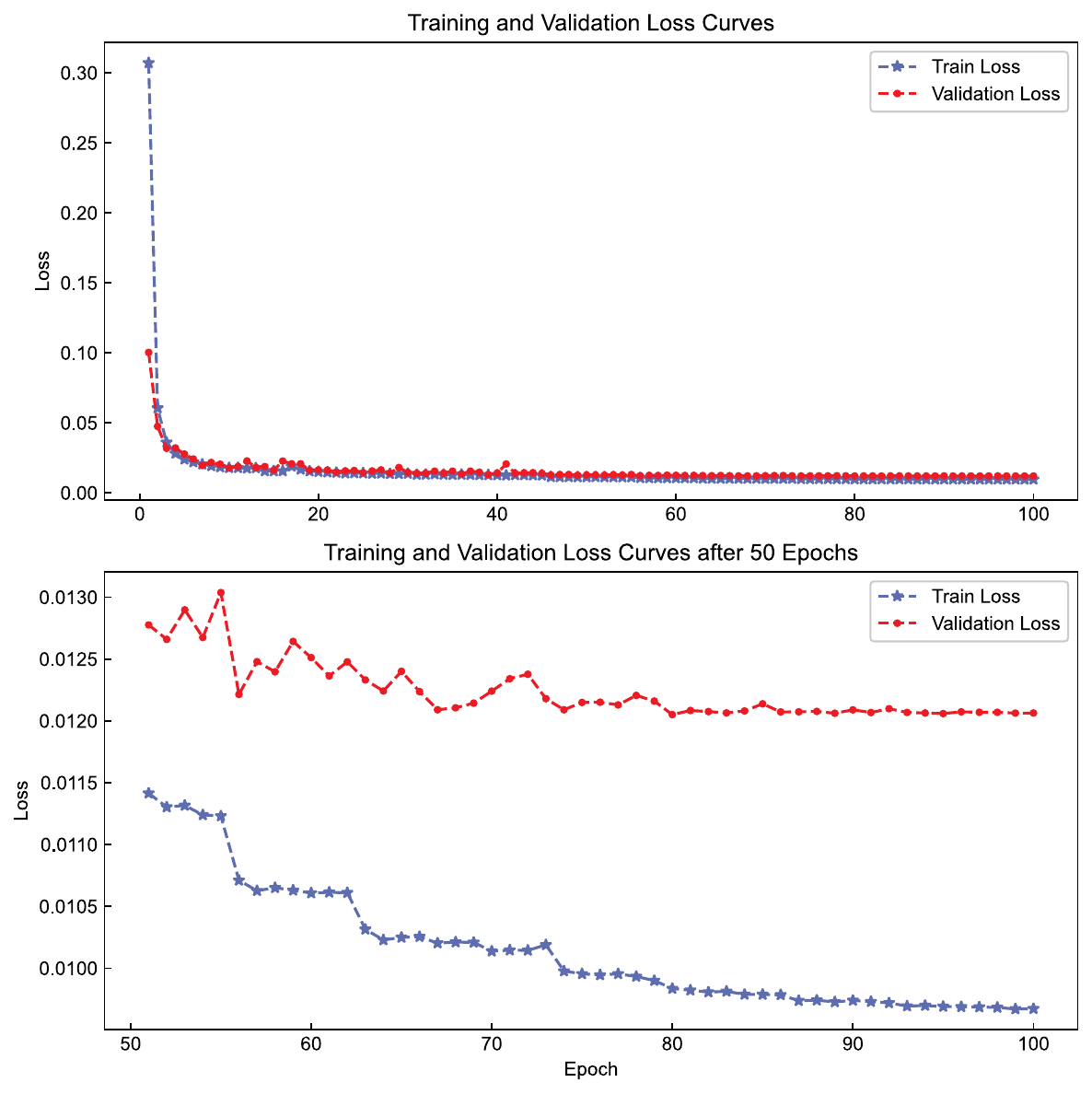}
    \caption{Training loss curve of the SageNet model.
    The model initially produces random predictions and learns rapidly within the first 10 epochs (\emph{upper panel}).
    After the 10th epoch, the learning curve gradually stabilizes, though the loss continues to decrease.
    By the 80th epoch (\emph{lower panel}), the validation loss curve flattens out at $\sim0.01$.} 
    \label{fig:train loss}
\end{figure}

In our experiments, the network is trained using the MSE loss function with the AdamW optimizer,
configured with a learning rate of $\alpha = 10^{-4}$
and weight decay of $\lambda = 10^{-5}$.
To improve robustness, we inject randomness
into the neural network weight initialization
and the batch construction during data loading.
Our implementation utilizes \verb|PyTorch| \citep{paszke_pytorch_2019} for GPU acceleration,
achieving convergence in approximately 3 hours
on a single NVIDIA RTX 4090 GPU with 24 GB of memory.
The train and validation loss curves, shown in Fig.~\ref{fig:train loss},
converge relatively quickly, reducing the need for extended training durations.

\section{Performance Analysis}\label{sec:performance}

\subsection{Metrics}

To compare the performance of \verb|SageNet|
with the existing numerical method in the \verb|stiffGWpy| code,
we reserve 10\% of the data set for testing.
Since the length of the GW spectrum curve varies among the samples,
we adopt the Symmetric Mean Absolute Percentage Error (SMAPE) as our error metric.
SMAPE assesses the relative error magnitude
by averaging the symmetric absolute percentage differences between predicted and actual values:
\begin{equation}
    \mathrm{SMAPE} = \frac{1}{256} \sum_{i=1}^{256} \frac{2 |A_i - F_i|}{|A_i| + |F_i| + 10^{-10}} \times 100\%.
\end{equation}
To avoid division by zero, a small positive number ($10^{-10}$) is added to the denominator.

We evaluate the performance of the emulator
using both $\{f_i\}$ and $\{\log_{10}\Omega_{\mathrm{GW}}(f_i)\}$ as metrics,
combining their individual SMAPE values:
\begin{IEEEeqnarray}{rl}\label{eq:SMAPE_eval}
    \mathrm{SMAPE}_{\mathrm{Eval}}
    = & \,\frac{1}{256} \sum_{i=1}^{256} \left[\mathrm{SMAPE}_{f_i} 
    + \mathrm{SMAPE}_{\log_{10}\Omega_\mathrm{GW}(f_i)}\right] \nonumber\\
    & \,\times \,100\%.
\end{IEEEeqnarray}
This is our main approach to assessing the emulator's performance in the paper.

Alternatively, we can also construct an observation-driven SMAPE
which is not subject to the artificial frequency sampling pattern of $\{f_i\}$,
by interpolating the GW spectra at fixed frequencies inside the PTA and LIGO frequency bands. 
The PTA (nanoHertz) frequency band is interpolated at the 14 lowest frequencies
from the NANOGrav 15 yr data set \citep{2023ApJ...951L...8A}. 
The LIGO band, spanning 20 to 1726\,{Hz}, is interpolated at uniform intervals
in accordance with the LIGO frequency resolution \citep{2021PhRvD.104b2004A}.
We compute the mean absolute errors 
between the interpolated values on the predicted spectrum and those on the true spectrum,
$\log_{10}\Omega_{\mathrm{GW}}(f_j)$, and then construct the following SMAPE:
\begin{equation}\label{eq:SMAPE_GW}
    \mathrm{SMAPE}_{\Omega_{\mathrm{GW}}}
    = \frac{1}{m} \sum_{j=1}^m \mathrm{SMAPE}_{\log_{10}\Omega_{\mathrm{GW}}(f_j)} \times 100\%,
\end{equation}
where $m$ is the number of the interpolated points on $\log_{10}\Omega_{\mathrm{GW}}(f)$,
and $f_j$ denotes the above designated frequencies inside the PTA and LIGO bands.
This approach provides a robust evaluation of the emulator's performance pertinent to the current SGWB measurements.

\begin{figure}[ht!]
	\centering
	\subfigure{\includegraphics[width=.45\textwidth]{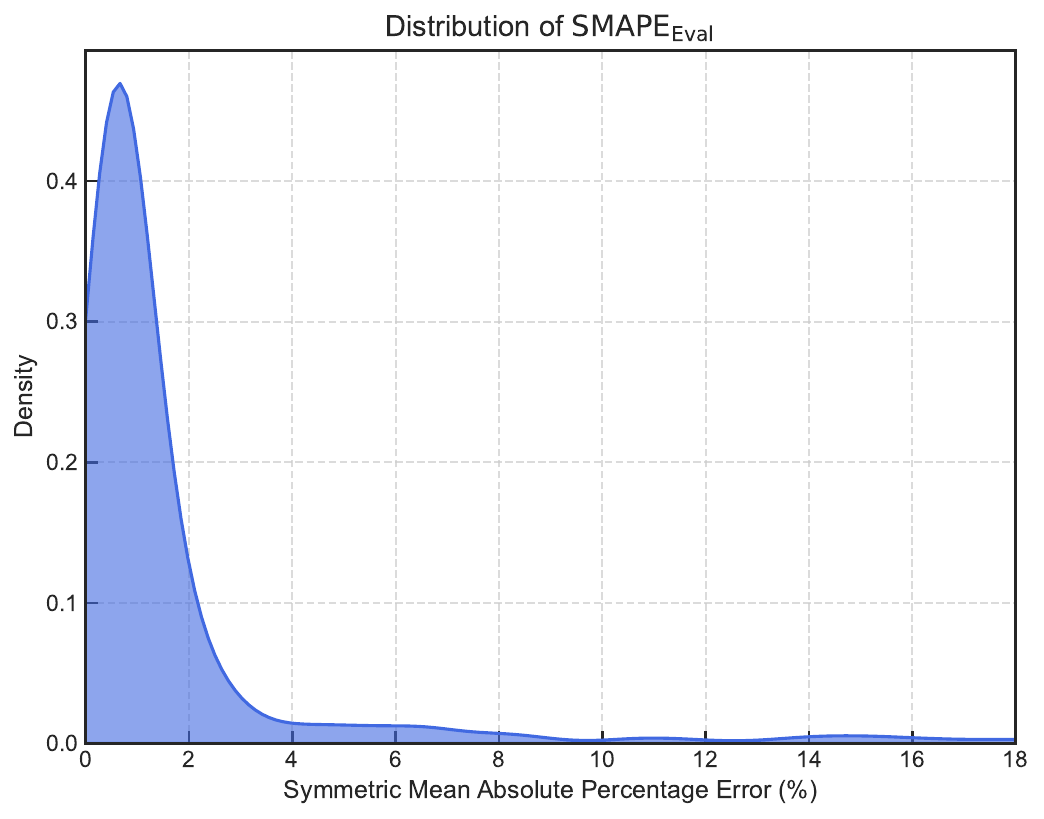}}
    \caption{Error distribution of test emulations.
    This histogram illustrates the performance of the emulator predictions
    against the ground truth of the test samples.
    The horizontal axis represents the dimensionless SMAPE
    and the vertical axis represents the normalized frequency counts (bin areas standardized to 1).
    Notably, 90.9\% of predictions achieve a $\mathrm{SMAPE}_\mathrm{Eval}$ of under 4\%
    and 86.0\% of the samples have an $\mathrm{SMAPE}_\mathrm{Eval}$ of less than 3\%,
    highlighting the accuracy and robustness of our emulator.
    A small fraction of the predicted samples exhibits higher SMAPE values,
    attributed to the fact that the predicted $\{f_i\}$ sequence
    is not necessarily in ascending order as is the true GW spectrum sequence.
    These errors are hence not real and can be eliminated
    by reordering the predicted $\{f_i,\,\log_{10}\Omega_\mathrm{GW}(f_i)\}$ sequence.}
    \label{fig:error distribution}
\end{figure}

\begin{figure*}[ht!]
    \centering
    \includegraphics[width=0.85\linewidth]{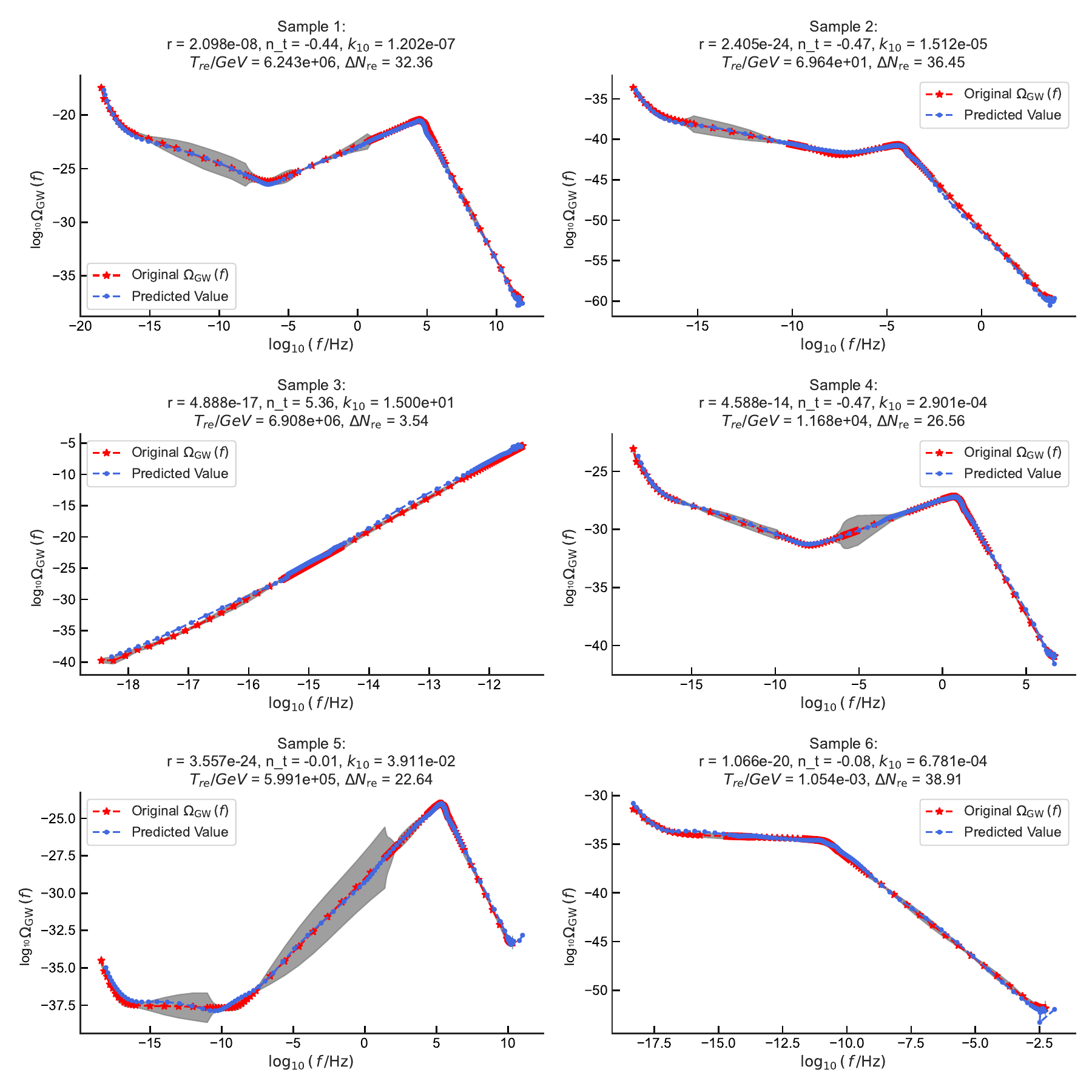}
    \caption{Example emulation results. 
    Each panel corresponds to a distinct type of the GW spectra.
    The emulator consistently predicts the correct evolutionary trends, variations, frequency limits,
    and specific $\log_{10}\Omega_{\mathrm{GW}}(f_i)$ values,
    regardless of the variability in the spectral shapes.
    Surprisingly, it also reproduces the density distribution of the sample points along the frequency axis,
    accurately reconstructing the sampling patterns in both high-variation and quasi-linear segments of the curves.
    The gray areas represent the predicted error distribution,
    for which the errors in both $f_i$ and $\log_{10}\Omega_{\mathrm{GW}}(f_i)$ are combined
    to obtain a more rigorous assessment than standard methods.
    In sparsely sampled segments, although the positions of the predicted points (blue)
    deviate from the original data (red),
    these points lie on almost the same curve as the original ones.
    }\label{fig:sample testing}
\end{figure*}

\begin{figure*}
    \centering
    \includegraphics[width=0.85\linewidth]{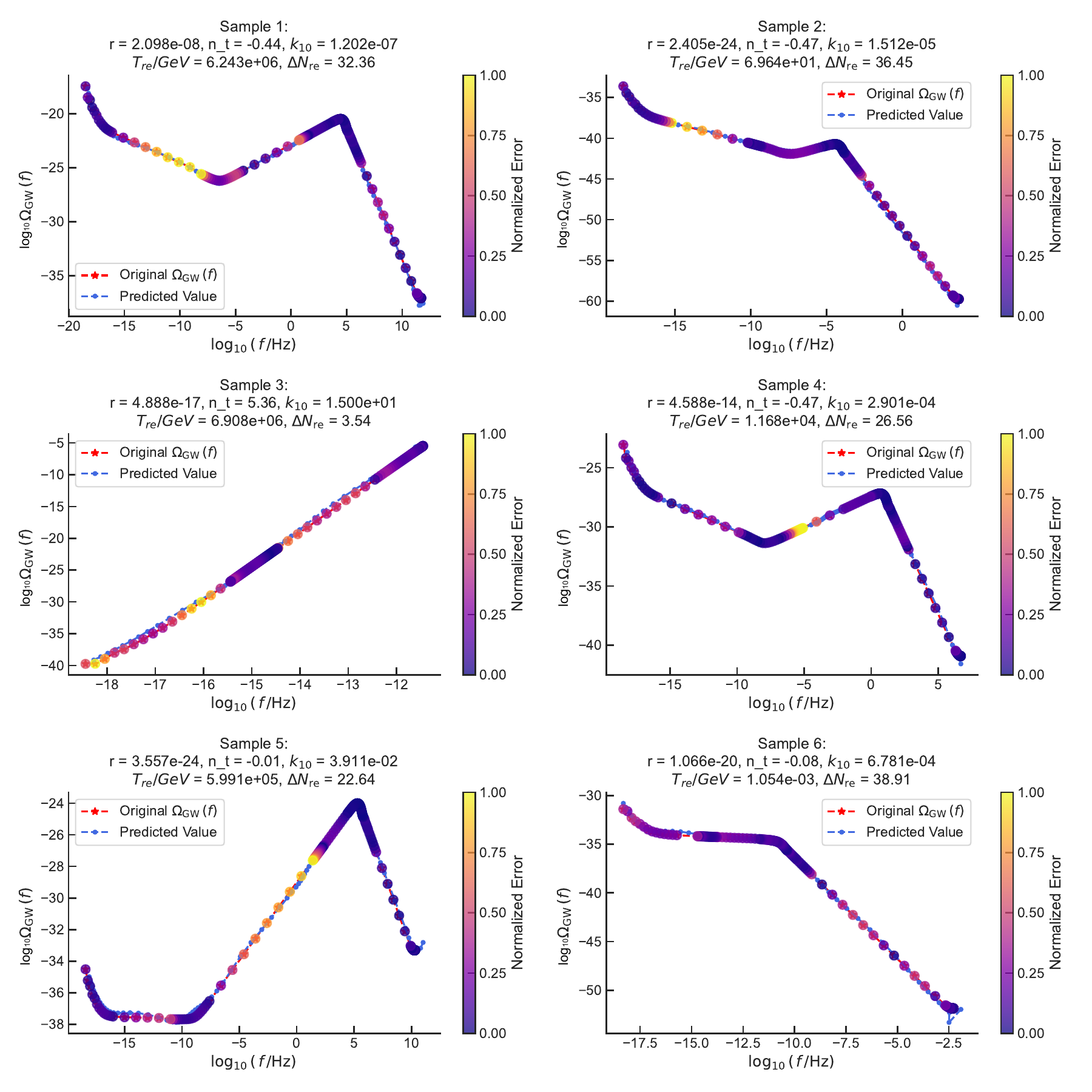}
    \caption{Error density distribution across predicted points.
    Errors are normalized to the interval $[0,1]$ and color-coded.
    High-error points (yellow) are concentrated in sparsely sample segments of the curves.
    Nevertheless, these errors do not undermine the overall accuracy
    of the GW spectra reconstructed by SageNet.
    }\label{fig:sample testing density}
\end{figure*}

\begin{figure*}
    \centering
    \includegraphics[width=0.92\linewidth]{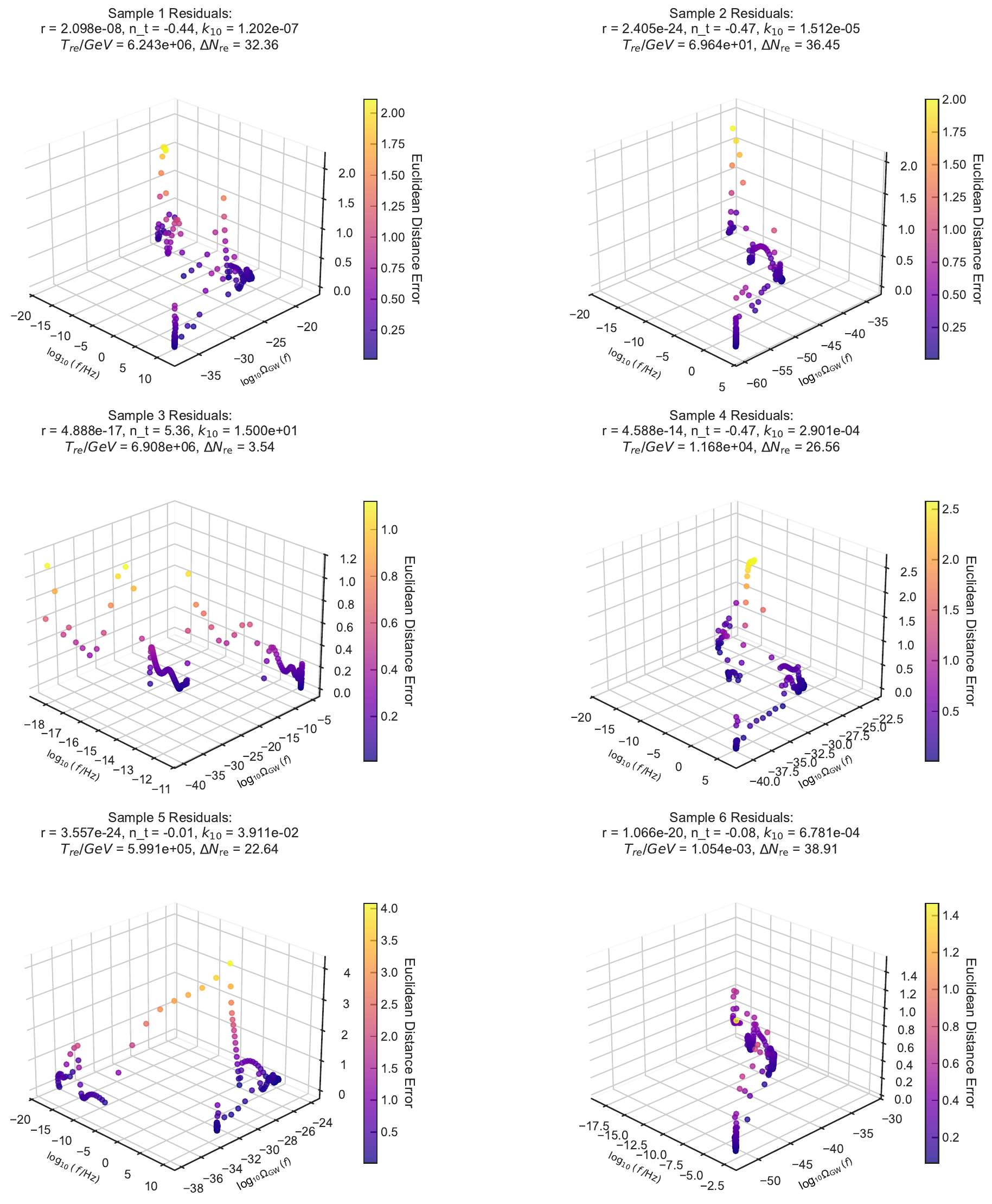}
    \caption{Three-dimensional error density distribution.
    This visualization highlights that larger errors occur at sparsely sampled segments of the curves,
    while the densely sampled segments exhibit minimal residuals.
    }\label{fig:sample testing 3d}
\end{figure*}

\begin{figure}[ht!]
	\centering
	\subfigure[Distribution of the original samples.] {\includegraphics[width=.48\textwidth]{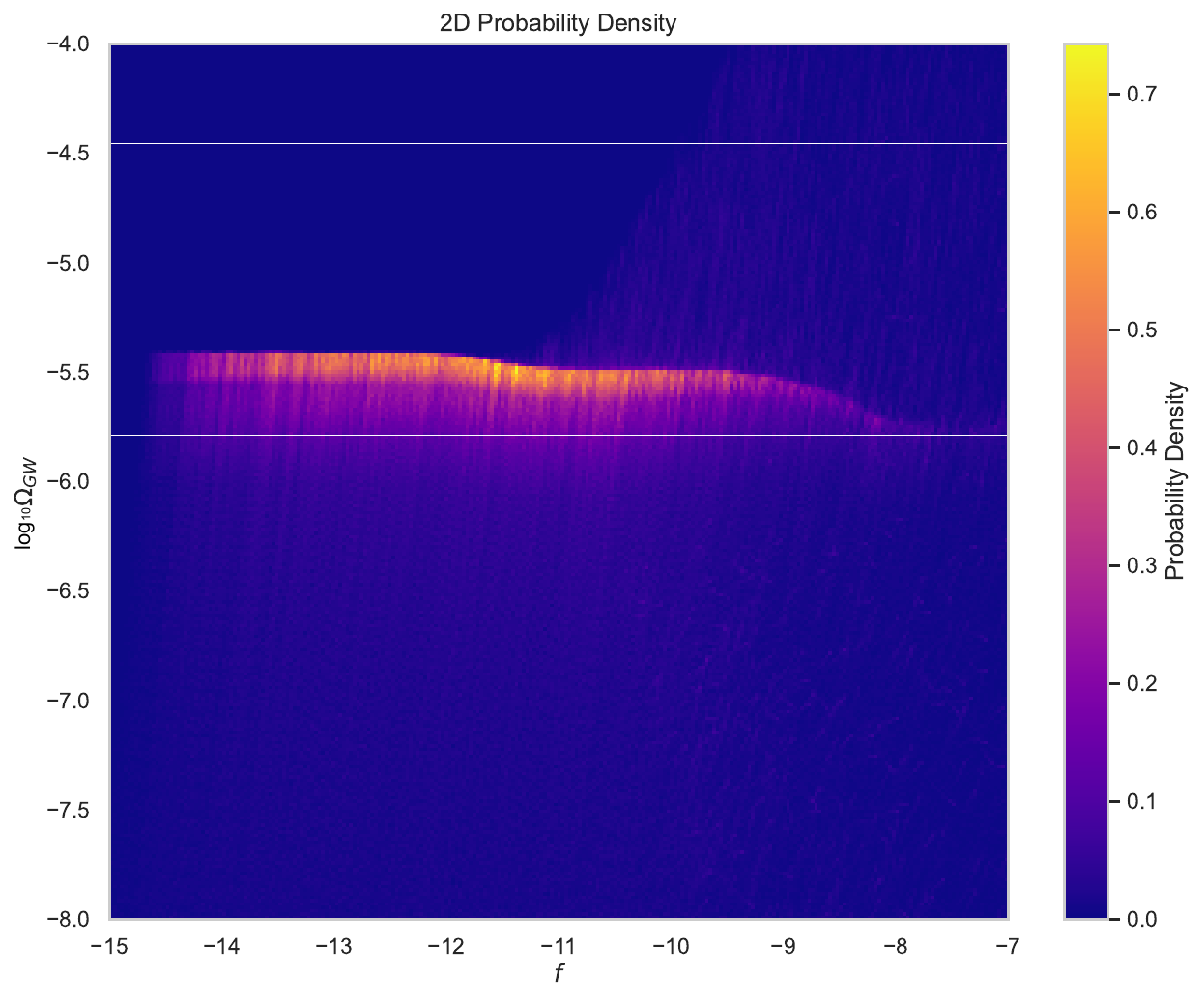}}
	\subfigure[Distribution of the predicted samples.] {\includegraphics[width=.48\textwidth]{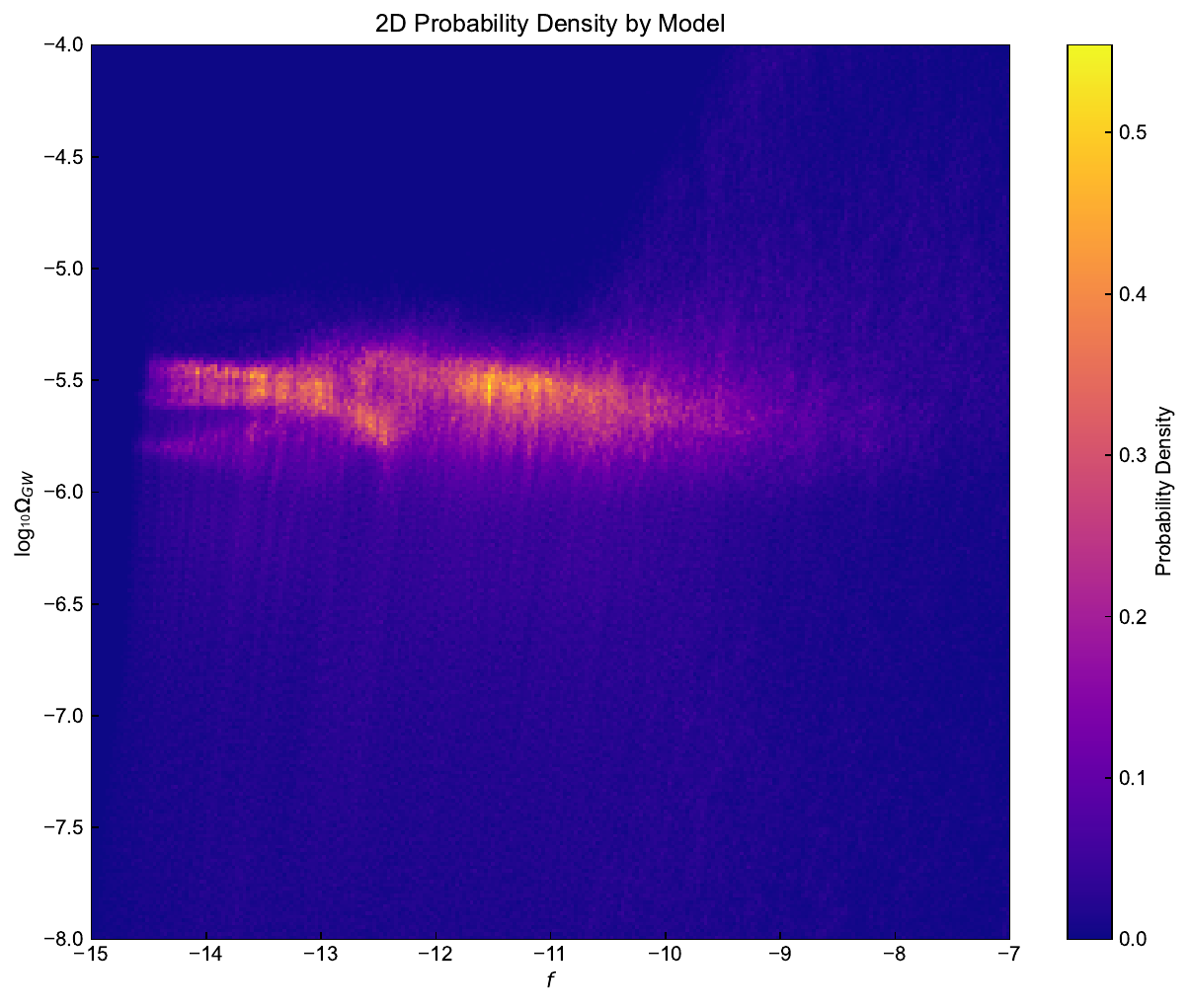}}
    \caption{Comparison between the distribution of the original numerical samples
    and that of the predicted samples from emulations.
    The apparent resemblance demonstrates that SageNet can learn both physical information
    and artificial sampling patterns simultaneously.}
    \label{fig:sampling learning}
\end{figure}

\subsection{Main Accuracy Evaluation: $\mathrm{SMAPE}_\mathrm{Eval}$}

The error distribution of the test results,
shown in Fig.~\ref{fig:error distribution},
indicates that 90.9\% (86.0\%) of the predictions for $\Omega_\mathrm{GW}(f)$
can achieve a SMAPE of less than 4\% (3\%). 
We report these numbers as part of the main results of the paper.
To further illustrate the performance of \verb|SageNet|,
we present random example emulations shown in Fig.~\ref{fig:sample testing}.
The predicted GW spectra exhibit good agreement with the respective ground truth curves,
accurately capturing key features such as
the amplitudes, the spectral indices
and the frequencies where the slopes change.
With a fixed sampling of 256 points per curve,
most errors arise from the shift of each predicted frequency,
$f_i$, along the spectrum curve,
rather than any shift away from the curve.

Taking advantage of the sequential nature
of the data points $\{f_i,\,\log_{10}\Omega_\mathrm{GW}(f_i)\}$,
our LSTM-based emulator can effectively leverage information
from adjacent points in \emph{densely} sampled segments of the curves,
thereby obtaining accurate predictions with minimal errors.
However, the emulator encounters challenges
in predicting points within \emph{sparsely} sampled segments,
leading to increased errors.
Still, the predicted curves (as shown in Fig.~\ref{fig:sample testing})
retain remarkable alignment with the ground truth
despite the error margins (gray areas).
This demonstrates the inherent fidelity of \verb|SageNet|.

The relationship between errors and
the sampling density along the $\Omega_\mathrm{GW}(f)$ curves
is further visualized in Fig.~\ref{fig:sample testing density},
where the normalized error distribution across the $[0,1]$ interval
is indicated by the color bar.
The color coding reveals that the points with the highest errors (yellow points)
lie on the sparsely sampled, monotonic, and nearly linear segments of the GW spectra.
It turns out that these localized errors do not
compromise the global accuracy of the curve reconstruction,
demonstrating the ability of \verb|SageNet| to accurately reconstruct
the present-day spectra of the inflationary SGWB,
even for sparsely sampled frequency ranges.
To reaffirm our conclusion,
we examine the three-dimensional visualization of the error distribution
along the reconstructed $\log_{10}\Omega_\mathrm{GW}(f)$ curves
in Fig.~\ref{fig:sample testing 3d}.
It confirms that the densely sampled segments of the curves
consistently yield the optimal emulation results.

Surprisingly, both Figs.~\ref{fig:sample testing density} \& \ref{fig:sample testing 3d}
show that \verb|SageNet| is also capable of learning the artificial information
of the adaptive sampling method implemented in \verb|stiffGWpy|.
This capability is further illustrated in Fig.~\ref{fig:sampling learning},
where the upper panel presents the distribution of the original data set
generated by our numerical method,
and the lower panel shows the distribution of
the predicted samples produced by \verb|SageNet|
with the same physical parameters as in the original samples.
These two distributions are nearly identical,
indicating that the emulator is able to learn not only
the GW spectra (physical information) from the training samples
but also the artificially introduced sampling patterns.
This dual learning capability establishes \verb|SageNet|
as a robust alternative to traditional numerical methods
for capturing key ``geometric'' information of the GW spectra,
i.e., where the spectral indices change.

\subsection{Alternative Accuracy Evaluation: $\mathrm{SMAPE}_{\Omega_{\mathrm{GW}}}$}

The above results show that
the main component of $\mathrm{SMAPE}_\mathrm{Eval}$ defined in Eq.~(\ref{eq:SMAPE_eval})
is the errors of the $\{f_i\}$ predictions.
Since the latter can, in principle, be reduced
by increasing the density of frequency sampling, 
it may be more informative to consider the errors of the $\Omega_\mathrm{GW}(f)$ predictions only.
Therefore, we here also assess the accuracy of the \verb|SageNet|
emulations using the alternative $\mathrm{SMAPE}_{\Omega_{\mathrm{GW}}}$ defined in Eq.~(\ref{eq:SMAPE_GW}).

Our results, detailed in Table~\ref{tab:smape_results},
demonstrate that \verb|SageNet| achieves high accuracy
in both the PTA and LIGO frequency bands.
In the PTA band, the average SMAPE is 1.30\%,
with 90\% of the samples having an SMAPE below 2.69\% and 93.18\% below 3\%.
In the LIGO band, the average SMAPE is 1.48\%,
with 90\% of the samples below 3.35\% and 93.44\% below 4\%.
These findings confirm that \verb|SageNet|
maintains excellent performance for designated frequencies related to SGWB observations,
proving it to be a reliable tool for reconstructing the present-day spectra of the inflationary SGWB.

\begin{table}[htbp]
    \centering
    \caption{SMAPE Statistics for PTA and LIGO Frequency Bands. The table above marks the quantile error values of samples in various proportions (25\%, 50\%, 75\%, 90\%). The following table shows the proportion of samples with errors less than each level (1\%, 2\%, 3\%, 4\%, 5\%, 10\%).}
    \label{tab:smape_results}
    \small
    \setlength{\tabcolsep}{1.5pt}
    \renewcommand{\arraystretch}{1.1}
    \begin{tabular}{lcc}
        \toprule
        \textbf{Metric} & \textbf{PTA} & \textbf{LIGO} \\
        \midrule
        \textbf{Quantiles} & & \\
        25\% & 0.52\% & 0.54\% \\
        50\% & 1.04\% & 1.02\% \\
        75\% & 1.82\% & 1.90\% \\
        90\% & 2.69\% & 3.35\% \\
        \midrule
        \textbf{Sample below thresholds} & & \\
        SMAPE in 1\% & 48.19\% & 49.15\% \\
        2\% & 79.31\% & 76.66\% \\
        3\% & 93.18\% & 87.23\% \\
        4\% & 98.04\% & 93.44\% \\
        5\% & 99.27\% & 96.30\% \\
        10\% & 100.00\% & 99.85\% \\
        \midrule
        \textbf{Average SMAPE} & 1.30\% & 1.48\% \\
        \bottomrule
    \end{tabular}
\end{table}

\subsection{Computational Efficiency}

Conventional numerical modeling
of the inflationary SGWB, as implemented in the \verb|stiffGWpy| code,
requires solving the tensor wave equation
for multiple intermediate variables iteratively.
In each iteration, the solver integrates a set of ODEs
for given initial conditions and time intervals
using the \verb|SciPy| package \citep{scipy}
to determine the final state $\{t_{\mathrm{max}}, \mathbf{y}({t_{\mathrm{max}}})\}$.
As noted previously, typical counts of sampled points/frequencies
fall within $[110,125]$ or $[200,256]$ in \verb|stiffGWpy|,
demanding hundreds of numerical solutions for a single run
and taking up to tens of seconds.
In addition, the sequential nature of the required iterations
prevents GPU acceleration.

\begin{figure}[ht!]
    \centering
    \includegraphics[width=1\linewidth]{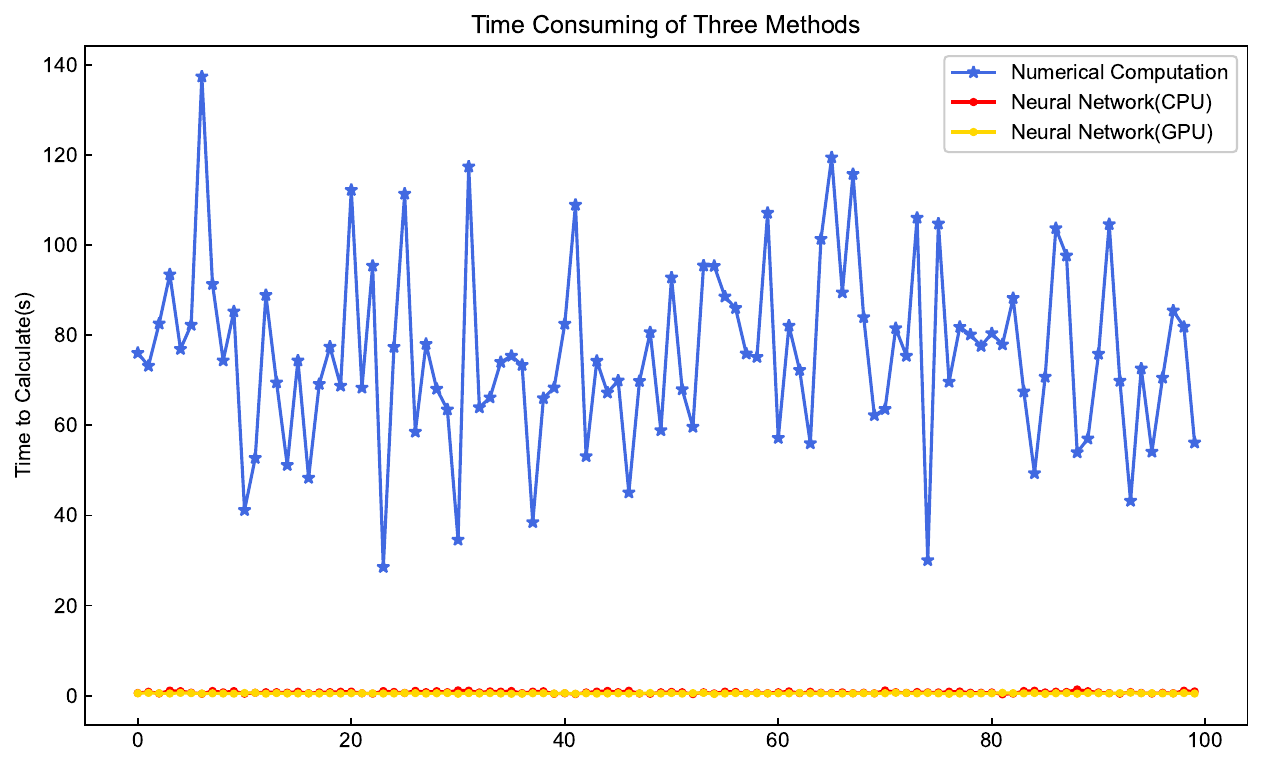}
    \caption{Wall times for producing 100 random samples using three methods.
    Brute-force numerical method shows significant variabilities in the computation times,
    for which even low-demand cases ($<100$ points) take $\mathcal{O}(10)$ seconds.
    By contrast, our neural network approach can output results
    within $\sim 7$\,ms on an Intel Core i7-11800H CPU,
    with the potential for further GPU acceleration.
    } \label{fig:time distribution}
\end{figure}

By contrast, our neural network emulator, \verb|SageNet|,
achieves an average inference time of 7.34 ms
on an Intel Core i7-11800H CPU
(see the time distribution in Fig.~\ref{fig:time distribution}),
representing a 10,000-fold speedup.
This acceleration is consistent with our expectations,
since neural networks replace iterative computations by direct nonlinear mappings.
The speed advantage of \verb|SageNet| could be further improved by GPU deployment,
leveraging mature GPU optimization frameworks.

Furthermore, brute-force numerical computations
exhibit significant temporal variabilities,
with durations ranging widely between cases,
as shown in Fig.~\ref{fig:time distribution}.
Even in minimally demanding scenarios ($<100$ sampled frequencies),
a single run typically requires $\mathcal{O}(10)$ seconds.
By contrast, our neural network approach, which eliminates iterative procedures,
achieves a stable computation time
of $\sim 0.007$ seconds (7\,ms) on an Intel Core i7-11800H CPU.
The above benchmarks based on this mainstream laptop processor
suggest that even greater efficiency gains are possible
with high-performance CPU/GPU architectures,
highlighting the potential of neural networks
to optimize temporal and spatial complexity.

In summary, we demonstrate that the neural network emulator
can achieve a significant acceleration in processing time,
compared with conventional numerical solvers.
The explanation is straightforward: during the training process,
it learns the solution patterns from extensive training data,
enabling a single rapid forward pass through the input.
As a result, \verb|SageNet| substantially reduces computational complexity
compared with traditional iterative schemes.
Once trained, the network can make robust predictions
of the present-day spectra of the stiff-amplified inflationary SGWB,
$\Omega_\mathrm{GW}(f)$, for arbitrary physical parameters
without performing time-consuming numerical calculations.

\section{Conclusion}\label{sec:conclusion}

We have explored the use of deep learning methods
to solve the coupled ODEs that govern the evolution of primordial tensor fluctuations,
which constitute the inflationary SGWB today.
Because of the backreaction of the SGWB on the expansion history,
exact and iterative integrations of these ODEs are necessary
for accurate predictions of the present-day GW spectrum.
However, this approach is computationally inefficient.
In this paper, we have shown that neural networks
can effectively approximate the energy density spectrum
of the inflationary SGWB with possible stiff amplification,
thereby significantly accelerating computation without sacrificing accuracy.

We presented \verb|SageNet|, an LSTM-based neural network
trained on data sets from numerical solutions of GW spectra.
The network allows accurate reconstructions of the spectra
and generalizes well to diverse cosmological parameters.
It successfully captures the important curve features
such as the spectral indices, the UV cutoff frequency,
and changes in indices due to cosmological transitions.
In addition, \verb|SageNet| is able to learn and reproduce
the artificial, adaptive sampling patterns implemented in the numerical calculations,
where the sampling of GW spectra is denser around changes of spectral indices
and more sparse in segments along simple power laws.
In summary, the neural network approach can yield reliable predictions
even without explicit feature engineering,
demonstrating its robustness for modeling the inflationary SGWB.

We performed a comparative analysis
between the conventional numerical approach (\verb|stiffGWpy|)
and the deep learning approach (\verb|SageNet|).
The results showed substantial efficiency gains;
the average execution time is reduced from tens of seconds in numerical methods
to milliseconds via neural network inference.
Therefore, neural network emulations of the stiff-amplified inflationary SGWB
enable Bayesian inferences on related extended cosmological models
using large GW and cosmological data sets.
From a broader perspective, our \verb|SageNet| framework
provides a fast, accurate, and generalizable solution
to the prediction of cosmological observables
that typically involves costly differential equation solvers.

\begin{acknowledgments}
FZ would like to thank the MIT LIGO Laboratory for its continuous support and advice.
FZ is supported by Ministry of Science and Technology of the People's Republic of China
(Grant No. 2023ZD0120704 of Project No. 2023ZD0120700)
and the National Natural Science Foundation of China (Grant No. 62372409).
BL is supported by the National Natural Science Foundation of China
(Grant Nos. 12203012, 12494575) 
and Guangxi Natural Science Foundation (Grant No. 2023GXNSFBA026114).
This work is also supported by the Guangxi Talent Program
(``Highland of Innovation Talents'').
JM is supported by the US~Department of Energy under Grant~\mbox{DE-SC0010129} and by NASA through Grant~\mbox{80NSSC24K0665}.
PRS acknowledges support from NASA under Grant No. 80NSSC22K175.
\end{acknowledgments}

\software{SageNet (\url{https://github.com/YifangLuo/SageNet}),
	       stiffGWpy (\url{https://github.com/bohuarolandli/stiffGWpy}),
	       PyTorch \citep{paszke_pytorch_2019}
          }


\appendix

\section{Details of the Model} \label{app:model}

In this appendix, we provide some supplementary details
of the physical model described in Section~\ref{sec:SGWB},
as well as some technical details of the numerical algorithm
implemented in the \verb|stiffGWpy| code.

\subsection{Thermal History}\label{app:thermal}

\begin{figure*}[ht!]
    \resizebox{\textwidth}{!}{\includegraphics{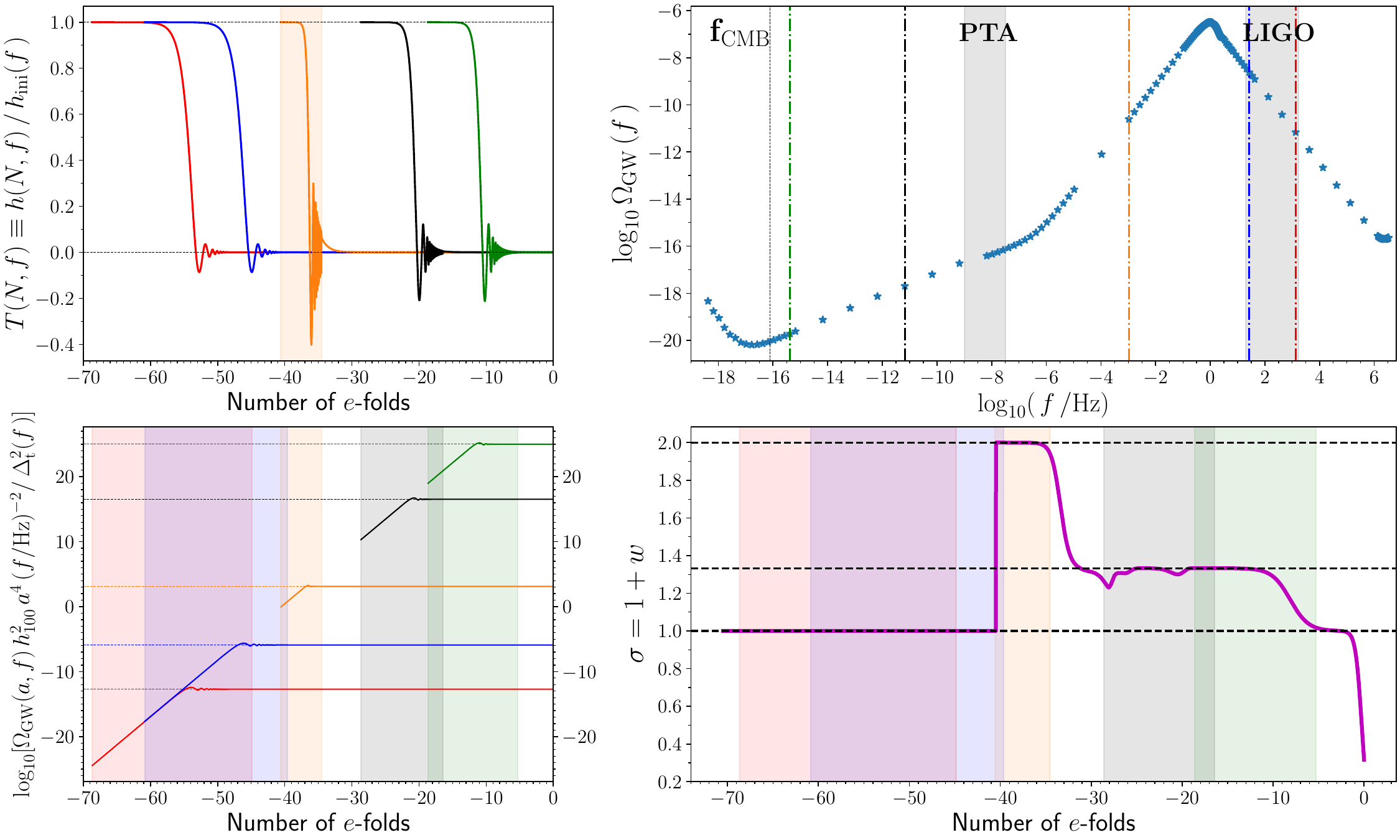}}
    \caption{Similar to Fig.~\ref{fig:solution},
    except that the \emph{lower left} panel displays the variable
    defined in Eq.~(\ref{eq:asymtoptic}).
    The vertical shades in the \emph{upper right} panel
    indicate the sensitive frequency band of PTAs
    and that of Advanced LIGO/Virgo \citep{2021PhRvD.104b2004A},
    respectively. The vertical shades in all other panels
    indicate the integration time spans of the illustrative modes. 
    } \label{fig:asymp}
\end{figure*}

By the end of reheating (at $T_\mathrm{re}$),
all the energy of the inflaton field has been transferred
to the stiff fluid and the Standard Model (SM) particles,
which marks the onset of the thermal history; see Fig.~\ref{fig:sigma}.
The energy density of the SM particles in the thermal bath
is usually parameterized by $g_{*}(T)$, 
the effective number of relativistic degrees of freedom
as a function of temperature, so that $\rho_\mathrm{SM}(T)=g_{*}(T)\,T^4$.
In the \verb|stiffGWpy| code \citep{2021JCAP...10..024L, 2025ApJ...985..117L},
the thermal history or the $g_{*}(T)$ function
is adopted from the tabulated function in \citet{2020JCAP...08..011S} for $T\geq10~\mathrm{MeV}$,
and computed using the \verb|FortEPiaNO| package
\citep{2019JCAP...07..014G} for $T<10~\mathrm{MeV}$,
which incorporates accurate prescriptions 
for the out-of-equilibrium neutrino decoupling \citep{2005NuPhB.729..221M}.
As shown in the lower right panel of Fig.~\ref{fig:asymp},
the earlier features around $N\sim-30$ are the results of the QCD phase transition.
The dent at $N\sim-20$ is due to electron-positron annihilation,
which overlaps neutrino decoupling.

\subsection{UV Cutoff in Primordial Tensor Spectrum}\label{app:cutoff}

Tensor perturbations generated by inflation
are naturally subject to a UV frequency cutoff,
$f_\mathrm{inf}$, corresponding to the horizon scale at the end of inflation.
However, when $n_\mathrm{t}>0$, 
the tensor amplitudes from the blue-tilted, power-law primordial spectrum
can reach unity at some high frequency \emph{below} the $f_\mathrm{inf}$ cutoff.
When such apparent nonlinearity happens, the tensor amplitude must saturate, 
and a more advanced prescription than linear perturbation theory
is required to treat the UV behavior of the primordial tensor fluctuations
\citep[e.g.,][]{2023MNRAS.520.1757G,2024JHEP...02..008Y,2024PhRvD.110j3529P}.
This potential pathology in the primordial tensor power spectrum
has been pointed out by \citet{2025ApJ...985..117L}.
Unfortunately, it is often overlooked in existing GW data analyses
based on the inflationary SGWB model. 

In the \verb|stiffGWpy| code,
we avoid dealing with the unknown physics beyond the apparent nonlinearity,
by imposing an additional UV cutoff on the primordial tensor spectrum,
$f_\mathrm{cut}$, at which the tensor power equals unity,
such that $A_\mathrm{t}\,(f_\mathrm{cut}/f_\mathrm{CMB})^{n_\mathrm{t}}\equiv 1$.
This cutoff is illustrated in the inset of the upper right panel of Fig.~\ref{fig:solution}.

\subsection{Integration of the Dynamical System}\label{app:numerical}

As mentioned in Section~\ref{sec:SGWB},
we solve the exact tensor wave equation
for a range of sampled frequencies, using a dynamical system approach.
For each frequency, the dynamical system (\ref{eq:zeta}–\ref{eq:y})
is integrated over a time interval that encompasses its horizon-crossing epoch,
from an initial state when $2\pi f \ll aH$ to a final state when $2\pi f \gg aH$.

The choice of this integration window is not unique.
In the implementation of \cite{2021JCAP...10..024L} and \verb|stiffGWpy|,
this time window begins at $2\pi f/aH = e^{\zeta_0} = 10^{-3}$
and ends at $\zeta_f = 5$ (where $2\pi f/aH = e^5 \approx 150$)
for all frequencies, as mentioned in Section~\ref{sec:design}.
This is a balanced choice between computational accuracy and efficiency.
On the other hand, the fixed interval of $[\zeta_0, \zeta_f]$
corresponds to different time spans in terms of $e$-folds for different modes,
as evidenced by the widths of the shaded regions in Fig.~\ref{fig:asymp}.

In the superhorizon limit, the inflationary tensor modes are frozen,
so that the initial conditions for the tensor transfer function are given by
$T \simeq 1$ and $\dot{T} \simeq 0$.
Consequently, we take $(\zeta_f, x_f, y_f)\simeq(\zeta_0, 0, e^{\zeta_0})$
for their superhorizon initial conditions, independent of frequency.
These conditions are depicted in the upper left panel of Fig.~\ref{fig:asymp}.

In the subhorizon limit, tensor modes are highly oscillatory
and we must switch to the time-averaged solution,
as demonstrated in the upper left panel of Fig.~\ref{fig:asymp}.
The resultant GWs redshift as radiation,
with energy density scaling as $\rho_\mathrm{GW} \propto a^{-4}$,
indicated by the lower left panel of Fig.~\ref{fig:asymp}.
This panel displays the following dimensionless variable
that converges to an asymptotic value for each tensor mode in the subhorizon limit:
\begin{equation}\label{eq:asymtoptic}
    \frac{\Omega_\mathrm{GW}(a,f)}{\Delta^2_\mathrm{t}(f)}\frac{H^2a^4}{f^2}
    \to\mathrm{const}=\frac{\Omega_\mathrm{GW}(f)\,H_0^2}{\Delta^2_\mathrm{t}(f)\,f^2},
    \qquad \mathrm{as}\quad\frac{2\pi f}{aH}\to +\infty.
\end{equation}
For convenience, we use $h_{100}\equiv H/(100\,\mathrm{km\,s}^{-1}\,\mathrm{Mpc}^{-1})$
instead of $H$ in the lower left panel of Fig.~\ref{fig:asymp}.
It shows that for each frequency, the convergence is already reached
by the end of its integration time span,
when the mode is well inside the horizon ($2\pi f/ aH=e^5$).


\bibliography{SGWB_emulator_updated}{}
\bibliographystyle{aasjournalv7}



\end{document}